\documentclass[]{tMPH2e}
\usepackage{xcolor}
\usepackage{tikz}
\definecolor{dgreen}{rgb}{0,.5,0}
\definecolor{dred}{rgb}{.7,.0,.0}
\def\etal{{\it et al.}}

\def\ddroit{{\rm d}}

\begin{document}
\doi{10.1080/0026897YYxxxxxxxx}
 \issn{}
\issnp{}
\jvol{00}
\jnum{00} \jyear{2013} 

\markboth{Odile Franck {\it et al.}}{Molecular Physics}

\articletype{Manuscript}

\title{{\itshape 
Generalized adiabatic connection in ensemble 
density-functional theory for excited states:
example of the H$_2$ molecule
}
}

\author{
Odile Franck and
Emmanuel Fromager$^{\ast}$\thanks{$^\ast$Corresponding author.
Email: fromagere@unistra.fr 
\vspace{6pt}}
\\\vspace{6pt}  
{\em{
Laboratoire de Chimie Quantique,
Institut de Chimie, CNRS / Universit\'{e} de Strasbourg,
4 rue Blaise Pascal, 67000 Strasbourg, France
}};\\\vspace{6pt}  
}

\maketitle

\begin{abstract}

A generalized adiabatic connection for ensembles (GACE) is presented. In
contrast to the traditional adiabatic connection formulation, both
ensemble weights and interaction strength can vary along a GACE path
while the ensemble density is held fixed. The theory is presented for
non-degenerate two-state ensembles but it can in principle be extended
to any ensemble of fractionally occupied excited states. Within such a
formalism an exact expression
for the ensemble exchange--correlation density-functional energy, in terms of
the conventional ground-state exchange--correlation energy, is obtained by
integration over the ensemble weight. Stringent constraints on the
functional are thus obtained when expanding the ensemble exchange--correlation
energy through
second order in the ensemble weight. For illustration purposes, the
analytical derivation of the GACE
is presented for the H$_2$ model system in a minimal
basis, leading thus to a simple density-functional approximation to the
ensemble exchange--correlation energy. Encouraging results were obtained
with this approximation for the description in a large basis of the
first $^1\Sigma^+_g$ excitation in 
H$_2$ upon bond stretching. Finally, a range-dependent GACE has been
derived, providing thus a pathway to the development of a rigorous
state-average multi-determinant density-functional theory. 

\bigskip

\begin{keywords}
Ensemble Density-Functional Theory,
Excited States,
Adiabatic Connection,
Multiple Excitations,
Range Separation
\end{keywords}\bigskip

\end{abstract}

\section{Introduction}\label{sec:intro}

Time-dependent density-functional theory (TD-DFT) has become over the years the method of choice
for modeling excited-state properties of electronic 
systems~\cite{Casida_tddft_review_2012} due to its
lower computational cost, relative to wavefunction-based
methods, and its relatively good accuracy. Nevertheless, standard
TD-DFT calculations rely on the adiabatic approximation and,
consequently, they cannot describe multiple excitations. Remedies
have been proposed to cure TD-DFT in that respect but their accuracy
usually lags behind {\it ab initio}
methods~\cite{Casida_tddft_review_2012}. 

Let us stress that, even though they are not as popular as TD-DFT,
alternative time-independent DFT approaches
for excited states
have
been investigated over the years at both
formal and computational
levels~\cite{JPC79_Theophilou_equi-ensembles,SSC82_Stoddart_exchange_for_ensemble,PRA86_Kohn_LDAfun_ensembles,PRA_GOK_EKSDFT,PRA_Levy_XE-N-N-1,PRA96_Goerling_DFT_ES,PRA99_Goerling_DFT_beyond_HK_th,PRL99_Nagy_DFT_individual_ES,PRA99_Nagy_EDFT_Koopmans_th,PRL02_Gross_spurious_int_EDFT,PRL03_Pan_quantal_DFT_degenerate_states,PRL04_Burke_lack_of_HK_th_ES,PRL04_Burke_lack_of_HK_th_ES_err,JPCA08_Filatov_SA-SREKS-DFT,PRA09_Ayers_Tind-DFT_ES,JCP09_Ziegler_relation_TD-DFT_VDFT,PRA12_Nagy_TinD-DFT_ES,JCTC13_Ziegler_SCF-CV-DFT}.
We shall focus in this paper on DFT for ensembles of
fractionally occupied excited states as formulated by
Gross, Oliveira and Kohn (GOK)~\cite{PRA_GOK_EKSDFT}. 
GOK-DFT relies on a Rayleigh--Ritz variational principle for
ensembles~\cite{PRA_GOK_RRprinc} which
generalizes the seminal work of
Theophilou~\cite{JPC79_Theophilou_equi-ensembles} on equi-ensembles. 
Pastorczak \etal~\cite{PRA13_Pernal_srEDFT} have recently shown that the Helmholtz
free-energy variational principle can be connected to the GOK
variational principle. Despite the substantial theoretical investigations of
ensemble DFT for excited states, GOK-DFT has been applied only to the calculation of
excitation energies in atoms and small
molecules~\cite{PRA_GOK_EKSDFT_He,CPL98_Nagy_apps_GOK-DFT,ParagiXCens,ParagiXens}. One of the reason for
the lack of success of GOK-DFT is the absence of appropriate
exchange--correlation functionals for ensembles. 

The adiabatic connection
(AC) formalism 
~\cite{LANGRETH:1975p1425,GUNNARSSON:1976p1781,Gunnarsson:1977,LANGRETH:1977p1780,SAVIN:2003p635}
has often been used as a guideline for the development of approximate 
ground-state exchange--correlation functionals and, as it became
recently possible to compute the AC for molecular systems using accurate
{\it ab initio}
methodologies~\cite{Teale:2009p2020,Teale:2010,Teale:2010b,AC_2blehybrids_Yann},
such a formalism 
could become effective in identifying and avoiding models that rely heavily on
error cancellations. One can naturally assume that this statement holds
also for ensemble exchange--correlation functionals. Indeed, Nagy~\cite{Nagy_ensAC} has shown
that the
ground-state AC formula for the exchange--correlation energy can be
easily extended to ensembles. Nevertheless, in
this formulation, the ensemble density that is held fixed along the AC path
depends on the ensemble weights. It then becomes difficult to
investigate, for a fixed density, the variation of the
exchange--correlation density-functional energy as the ensemble weights
vary. As shown by Gross~\etal~\cite{PRA_GOK_EKSDFT}, this
variation plays a crucial role in GOK-DFT. When computed for the
ground-state density, it corresponds to the exact deviation of the true
physical excitation energy from 
the  
energy gap between the Kohn-Sham (KS) lowest unoccupied (LUMO) and highest
occupied (HOMO) molecular orbitals.  
In
addition, a precise knowledge of the weight dependence of the ensemble
exchange--correlation energy for a fixed density would enable the
construction of density-functional approximations (DFAs) that rely on
conventional ground-state functionals. So far this has been investigated
semi-empirically~\cite{Nagy_enseXpot,ParagiXens,ParagiXCens}. 

In the light of these considerations, we propose in this work a
generalized AC for ensembles where the ensemble density
is held fixed along the AC path as both ensemble weights and interaction
strength vary. For clarity, the formalism is presented for non-degenerate two-state
ensembles but it can in principle be extended to any ensemble of
fractionally occupied excited 
states. 

The paper is organized as follows: exact AC formulae are first derived
and discussed in Sec.~\ref{sec:theory}. For illustration
purposes, ACs are then constructed analytically in
Sec.~\ref{subsec:theory:AC_min_basis} for the H$_2$ model system
in a minimal basis. A simple DFA is thus
obtained for two-state ensembles. This approximation is then tested
in Sec.~\ref{sec:discussion}
with a large basis and standard exchange--correlation functionals for the description of the first
$^1\Sigma^+_g$ excited state of H$_2$ upon bond stretching. As a
perspective and in connection with the recent work of Pastorczak
\etal~\cite{PRA13_Pernal_srEDFT}, we propose in Sec.~\ref{sec:EsrDFT} to construct a range-dependent
generalized AC, providing thus a pathway to the
development of a 
rigorous state-average multi-determinant DFT based on range separation.
Conclusions are given in Sec.~\ref{sec:conclusions}.

\section{Theory}\label{sec:theory}

Exact expressions for the exchange--correlation energy of non-degenerate
two-state ensembles
are investigated in this section. It is
organized as follows: after a short summary of the GOK-DFT approach
(Sec.~\ref{subsec:theory:Eksdft}) and a brief introduction to 
the AC formalism (Sec.~\ref{subsec:theory:ACNagyensemble}), a generalized AC, where both
weight and interaction strength can vary along the AC path while the
ensemble density is held fixed, is presented in
Sec.~\ref{subsec:theory:ACensemble}.     
An exact Taylor expansion for the ensemble exchange--correlation density
functional through
second order in the ensemble weight is thus derived 
and stringent constraints on the functional are obtained and analyzed 
in Sec.~\ref{subsec:stringent_constraints}. The construction of the generalized AC for ensembles is
finally discussed in Sec.~\ref{subsec:construct_gace}.

For pedagogical purposes, all adiabatic connections
will be derived as if the input density they rely on could be represented by 
non-, partially- and 
fully-interacting 
pure ground states 
as well as by 
non-, partially- and
fully-interacting 
non-degenerate two-state ensembles. 
The
Legendre--Fenchel-transform-based formalism
introduced in the following should however enable to tackle situations
where $v$-representability problems occur. This should obviously be
investigated further and is left for future work.    

Note also that the generalized AC discussed in this paper could possibly
be extended to ensembles of
near-degenerate or degenerate states for the purpose of representing ground-state densities 
of strongly
multi-configurational systems
or densities that are not pure-state-$v$-representable. Such situations
will not be discussed in details 
here. When a non-degenerate two-state ensemble is used for
representing a given density in the generalized AC we
propose, that density will be assumed to be also
pure-state-$v$-representable.  

\subsection{Gross--Oliveira--Kohn density-functional
theory}\label{subsec:theory:Eksdft}

Let $\tilde{\Psi}_1$ and $\tilde{\Psi}_2$
denote the ground and first excited states of an electronic system. Both fulfill the Schr\"{o}dinger equation 
\begin{eqnarray}\label{Schroeqtruesyst}
\Big(\hat{T}+\hat{W}_{\rm ee}+\hat{V}_{\rm ne}
\Big)\vert\tilde{\Psi}_i\rangle={E}_i\vert\tilde{\Psi}_i\rangle,
\hspace{0.8cm} i=1,2,
\end{eqnarray}
where $\hat{T}$ is the kinetic energy operator, $\hat{W}_{\rm ee}$
denotes the two-electron repulsion operator, and
$\hat{V}_{\rm ne}=\int \ddroit{\bf r}\; v_{\rm ne}({\bf
r})\,\hat{n}({\bf r})$ is the nuclear potential operator.
According to the GOK variational principle~\cite{PRA_GOK_RRprinc}, which generalizes the
seminal work of Theophilou~\cite{JPC79_Theophilou_equi-ensembles}, the following inequality
holds for any trial set of orthonormal wavefunctions $\Psi_1$ and
$\Psi_2$ and any weight in the range $0\leq w\leq \frac{1}{2}$:
\begin{eqnarray}\label{GOKvarprinciple_wf}
(1-w)\,\langle\Psi_1\vert\hat{T}+\hat{W}_{\rm ee}+\hat{V}_{\rm ne}\vert\Psi_1\rangle
+w\,\langle\Psi_2\vert\hat{T}+\hat{W}_{\rm
ee}+\hat{V}_{\rm ne}\vert\Psi_2\rangle\geq E^w,
\nonumber\\
\end{eqnarray}
where the lower bound is the exact ensemble energy
\begin{eqnarray}\label{ensembleener}
E^w=(1-w)\,E_1+w\,E_2.  
\end{eqnarray}
As shown by Gross~\etal~\cite{PRA_GOK_EKSDFT}, an important consequence of this variational principle is that the
ensemble energy is a functional of the ensemble density 
\begin{eqnarray}\label{ensembledens}
n^w({\bf r})=(1-w)\,n_{\tilde{\Psi}_1}({\bf
r})+w\,n_{\tilde{\Psi}_2}({\bf r}). 
\end{eqnarray}
The former can be determined variationally as follows
\begin{eqnarray}\label{ensenergyvarprinc}
E^w&=&\underset{n}{\rm min}\Big\{
F^w[n]+
\int \ddroit{\bf r}\,v_{\rm
ne}({\bf r})\,n({\bf r})
\Big\},
\end{eqnarray}
where the universal GOK functional, which is an
extension of the Hohenberg--Kohn (HK) functional~\cite{hktheo} to ensembles, can be
written as follows using a Levy--Lieb constrained-search formulation,
\begin{eqnarray}\label{Fwdef}
F^w[n]&=&\underset{\{\Psi_1,\Psi_2\}^w\rightarrow n}{\rm min}\Big\{
(1-w)\,\langle\Psi_1\vert\hat{T}+\hat{W}_{\rm ee}\vert\Psi_1\rangle
+w\,\langle\Psi_2\vert\hat{T}+\hat{W}_{\rm
ee}\vert\Psi_2\rangle\Big\}
\nonumber\\
&=&
(1-w)\,\langle\Psi^w_1[n]\vert\hat{T}+\hat{W}_{\rm ee}\vert\Psi^w_1[n]\rangle
+w\,\langle\Psi^w_2[n]\vert\hat{T}+\hat{W}_{\rm ee}\vert\Psi^w_2[n]\rangle.
\end{eqnarray}
The minimization in Eq.~(\ref{Fwdef}) is restricted to orthonormal sets
of wavefunctions $\{\Psi_1,\Psi_2\}^w$ whose ensemble density
$(1-w)\,n_{\Psi_1}+w\,n_{\Psi_2}$ equals $n$.
By analogy with KS-DFT, Gross~\etal~\cite{PRA_GOK_EKSDFT} proposed to split their
functional into a non-interacting kinetic energy contribution and a
complementary Hartree--exchange--correlation (Hxc) term, 
\begin{eqnarray}\label{fwsplitks}
F^w[n]&=&T^w_{\rm s}[n]+{E}^w_{\rm Hxc}[n],
\end{eqnarray}
where
\begin{eqnarray}\label{Twsdef}
T^w_{\rm s}[n]&=&\underset{\{\Psi_1,\Psi_2\}^w\rightarrow n}{\rm min}\Big\{
(1-w)\,\langle\Psi_1\vert\hat{T}\vert\Psi_1\rangle
+w\,\langle\Psi_2\vert\hat{T}\vert\Psi_2\rangle\Big\}
\nonumber\\
&=&(1-w)\,\langle\Phi^w_1
[n]\vert\hat{T}\vert\Phi^w_1[n]\rangle
+w\,\langle\Phi^w_2
[n]\vert\hat{T}\vert\Phi^w_2[n]\rangle
\end{eqnarray}
is expressed in terms of the non-interacting ground $\Phi^w_1[n]$ and
first excited $\Phi^w_2[n]$ GOK determinants whose ensemble
density equals $n$. According to Eq.~(\ref{ensenergyvarprinc}), the
exact ensemble energy is expressed within GOK-DFT as 
\begin{eqnarray}\label{ksenergyvarprinc}
E^w&=&\underset{n}{\rm min}\Big\{
T^w_{\rm s}[n]+{E}^w_{\rm Hxc}[n]
+\int \ddroit{\bf r}\,v_{\rm
ne}({\bf r})\,n({\bf r})
\Big\}
\nonumber\\
&=&
(1-w)\,\langle\tilde{\Phi}^w_1
\vert\hat{T}\vert\tilde{\Phi}^w_1\rangle
+w\,\langle\tilde{\Phi}^w_2
\vert\hat{T}\vert\tilde{\Phi}^w_2\rangle+{E}^w_{\rm Hxc}[n^w]
\nonumber\\
&&+\int \ddroit{\bf r}\,v_{\rm
ne}({\bf r})\,n^w({\bf r}),
\end{eqnarray}
where the GOK determinants reproducing the exact ensemble density $n^w$ 
fulfill the following
self-consistent equations~\cite{PRA_GOK_EKSDFT}:
\begin{eqnarray}\label{sceqensembleKS}
&&\Big(\hat{T}+\hat{V}_{\rm ne}+\hat{{V}}^w_{\rm
Hxc}[n^w]\Big)\vert\tilde{\Phi}^w_i\rangle=
\mathcal{E}^w_{{\rm s},i}
\vert\tilde{\Phi}^w_i\rangle,
\hspace{0.4cm} i=1,2,
\nonumber\\
\nonumber\\
&&{\displaystyle
\hat{{V}}^w_{\rm Hxc}[n]=
\int \ddroit{\mathbf r}\,\frac{\delta {E}^w_{\rm Hxc}}{\delta n({\bf
r})}[n]\,\hat{n}({\bf r}).
}
\end{eqnarray}
An exact extension of KS-DFT to excited states is thus formulated.
Let us stress that, for a fixed density $n$, the
ensemble Hxc density-functional energy ${E}^w_{\rm Hxc}[n]$ varies with the ensemble
weight $w$. As shown by Gross~\etal~\cite{PRA_GOK_EKSDFT} and
discussed further in the rest of the paper, the exact deviation of the
physical excitation energy $E_2-E_1$ from the KS HOMO-LUMO gap is directly
related to this weight dependence. Note that the interacting and
non-interacting wavefunctions decorated
with a "$\sim$" are those that enable to reproduce the ensemble density
$n^w$ of the physical system with local potential $v_{\rm
ne}$. This notation will also be used in the following for
partially-interacting wavefunctions. 
\subsection{
Adiabatic connection formula for ensembles}
\label{subsec:theory:ACNagyensemble}

As shown by Nagy~\cite{Nagy_ensAC}, an exact expression can be derived for the
ensemble Hxc energy within the AC formalism. By 
analogy with the ground-state
formulation~\cite{LANGRETH:1975p1425,GUNNARSSON:1976p1781,Gunnarsson:1977,LANGRETH:1977p1780,SAVIN:2003p635},
we introduce auxiliary equations based on
a partially-interacting system,
\begin{eqnarray}\label{nagy_acw}
\Big(\hat{T}+\lambda\hat{W}_{\rm ee}+\hat{V}^{\lambda}
\Big)\vert\Psi^{\lambda}_i\rangle=\mathcal{E}_i^{\lambda}\vert\Psi^{\lambda}_i\rangle,
\hspace{0.2cm} i=1,2,
\end{eqnarray}
where $\Psi^{\lambda}_1$ and $\Psi^{\lambda}_2$ are the ground and first
excited auxiliary states, respectively. The local potential operator 
$\hat{V}^{\lambda}=\int \ddroit{\bf r}\,v^{\lambda}({\bf r})\,\hat{n}({\bf
r})$ ensures that the density constraint
\begin{eqnarray}\label{nagy_ACwdensconstraints}
n^w({\bf r})&=&(1-w)\,n_{\Psi_1^{\lambda}}({\bf
r})+w\,n_{\Psi_2^{\lambda}}({\bf
r})
\end{eqnarray}
is fulfilled for any interaction strength in the range $0\leq
\lambda\leq 1$.
Note that, for $\lambda=1$, $v^{\lambda}({\bf r})$ equals the nuclear
potential $v_{\rm ne}({\bf r})$ and
the wavefunctions $\Psi^{\lambda}_i$ reduce to the physical ones
$\tilde{\Psi}_i$, while for $\lambda=0$, $v^{\lambda}({\bf
r})$ reduces to the GOK potential $v_{\rm ne}({\bf r})+\delta {E}^w_{\rm
Hxc}[n^w]/\delta n({\bf r})$ and the auxiliary wavefunctions $\Psi^{\lambda}_i$ become the GOK determinants
$\tilde{\Phi}^w_i$.

According to Eq.~(\ref{fwsplitks}), the ensemble Hxc energy can be expressed as
\begin{eqnarray}\label{nagy_eq:EHxc_aux_ens_energies_1}
{E}^w_{\rm Hxc}[n^w]&=&
F^w[n^w]-T^w_{\rm s}[n^w]
\nonumber\\
&=& 
\int^1_0\ddroit\lambda\,\frac{\ddroit F^{\lambda,w}[n^w]}{\ddroit\lambda}
,
\end{eqnarray}
where we introduced the partially-interacting GOK functional 
\begin{eqnarray}\label{nagy_Fxlambda_def}
F^{\lambda,w}[n^w]&=&
(1-w)\,\langle\Psi^{\lambda}_1\vert\hat{T}+\lambda\hat{W}_{\rm
ee}\vert\Psi^{\lambda}_1\rangle
+w\,\langle\Psi^{\lambda}_2\vert\hat{T}+\lambda\hat{W}_{\rm
ee}\vert\Psi^{\lambda}_2\rangle.
\end{eqnarray}
Since, according to the Hellmann--Feynman theorem and the density constraint in
Eq.~(\ref{nagy_ACwdensconstraints}),
\begin{eqnarray}\label{nagy_eq:HFth_simplifications}
\frac{\ddroit F^{\lambda,w}[n^w]}{\ddroit\lambda}
&=&
(1-w)\,
\langle\Psi_1^{\lambda}\vert\hat{W}_{\rm
ee}\vert\Psi_1^{\lambda}\rangle
+w\,
\langle\Psi_2^{\lambda}\vert\hat{W}_{\rm
ee}\vert\Psi_2^{\lambda}\rangle
,
\end{eqnarray}
we finally recover the expression of Nagy~\cite{Nagy_ensAC}: 
\begin{eqnarray}\label{nagy_ACHxcw}
\displaystyle
E^w_{\rm Hxc}[n^w]=
(1-w)\int^1_0
\ddroit\lambda\,
\langle\Psi_1^{\lambda}\vert\hat{W}_{\rm
ee}\vert\Psi_1^{\lambda}\rangle
+w\int^1_0
\ddroit\lambda\,\langle\Psi_2^{\lambda}\vert\hat{W}_{\rm
ee}\vert\Psi_2^{\lambda}\rangle.
\end{eqnarray}
This formulation is appealing as it would potentially enable the
accurate calculation of ensemble Hxc energies from {\it ab initio}
methods~\cite{Teale:2009p2020,Teale:2010,Teale:2010b}. Nevertheless, the computed energies would be obtained
for a given ensemble density $n^w$ that depends on the ensemble weight $w$. In
other words, Nagy's AC cannot be used straightforwardly for computing the 
Hxc density-functional energy as the ensemble weight varies while the
density is fixed. Being able to perform such a calculation is
highly desirable as it would enable to develop DFAs for ensembles based
on conventional ground-state DFAs.  
Constructing an AC where the density is held fixed as both interaction
strength and ensemble weight vary is appealing in this respect.  
\subsection{
Generalized adiabatic connection for ensembles
}\label{subsec:theory:ACensemble}

In order to investigate the weight dependence of the universal ensemble Hxc
density functional $E^w_{\rm Hxc}[n]$, we propose to construct a {\it generalized adiabatic
connection for ensembles} (GACE) which is based on the following  
auxiliary equations,
\begin{eqnarray}\label{acw}
\Big(\hat{T}+\lambda\hat{W}_{\rm ee}+\hat{V}^{\lambda,\xi}
\Big)\vert\Psi^{\lambda,\xi}_i\rangle=\mathcal{E}_i^{\lambda,\xi}\vert\Psi^{\lambda,\xi}_i\rangle,
\hspace{0.2cm} i=1,2,
\end{eqnarray}
where the local potential operator
$\hat{V}^{\lambda,\xi}=\int \ddroit{\bf r}\,v^{\lambda,\xi}({\bf
r})\,\hat{n}({\bf r})$ ensures that the density constraint
\begin{eqnarray}\label{ACwdensconstraints}
n({\bf r})&=&(1-\xi)\,n_{\Psi_1^{\lambda,\xi}}({\bf
r})+\xi\,n_{\Psi_2^{\lambda,\xi}}({\bf
r})
\end{eqnarray}
is fulfilled not only for all interaction strengths in the range $0\leq
\lambda\leq 1$, but also for {\it all ensemble weights} in the range
$0\leq \xi\leq w$. In the particular case where $\xi=w$ and $n=n^w$, the
GACE reduces to Nagy's AC~\cite{Nagy_ensAC}. Let us stress that, for any
physical ensemble density $n$, there is in principle no guarantee that the local potential
$v^{\lambda,\xi}$ exists for all $\lambda$ and $\xi$ values. This
so-called "$v$-representability problem" can be addressed formally when
using a Legendre--Fenchel-transform formalism as discussed further in
Sec.~\ref{subsec:construct_gace}. As mentioned previously, we will
assume for pedagogical purposes that the density $n$ is $v$-representable for all
$\lambda$ and $\xi$ values.   

\begin{figure}
\centering
{
\includegraphics[scale=0.5]{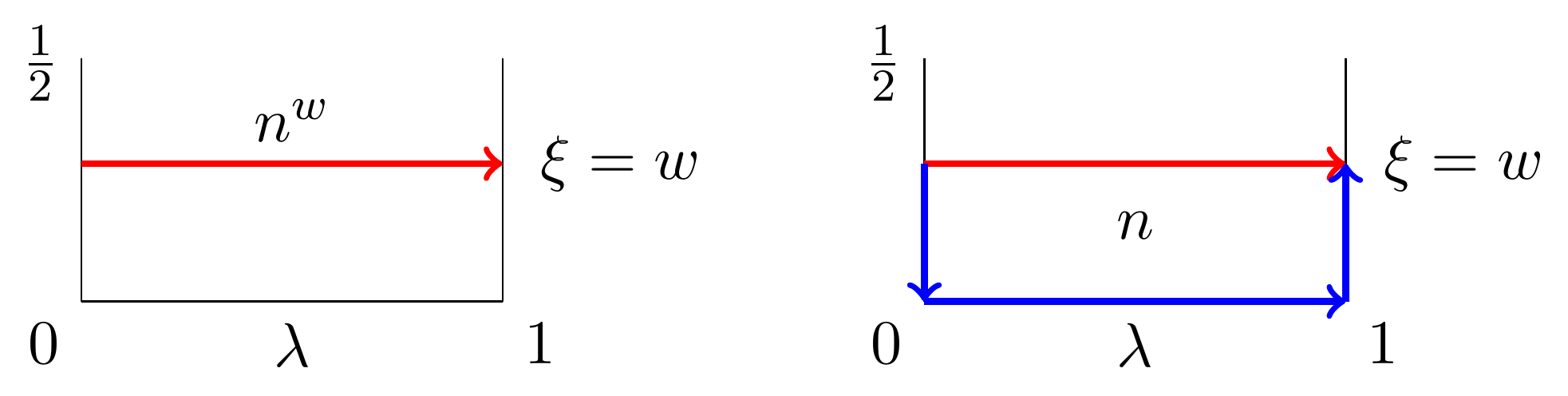}}
\caption{
Graphical representations of both traditional (left) and generalized
adiabatic connections (right) for a
two-state ensemble. See text for further details.
}\label{fig:ACandGACE}
\end{figure}

Since the ground-state Hartree density-functional energy expression is usually
employed for the ensemble Hartree energy~\cite{PRA_GOK_EKSDFT} 
\begin{eqnarray}\label{complementHensembledensity}
{E}^w_{\rm H}[n]&=&{E}_{\rm H}[n]=
\displaystyle \frac{1}{2}\int\int
\ddroit{\mathbf{r}}\ddroit{\mathbf{r'}}\frac{n(\mathbf{r})n(\mathbf{r'})}
{\vert {\bf r}-{\bf r'} \vert},
\end{eqnarray}
the latter is by definition weight-independent and the exact ensemble
exchange--correlation functional is defined as 
\begin{eqnarray}\label{complementxcensembledensity}
{E}^w_{\rm xc}[n]&=&{E}^w_{\rm Hxc}[n]-{E}_{\rm H}[n].
\end{eqnarray}
One of the advantage of the GACE
relative to Nagy's AC is that various adiabatic paths can be followed for
calculating the ensemble exchange--correlation energy. In order to
connect the ensemble exchange--correlation functional to its
ground-state ($w=0$) limit $E_{\rm xc}[n]$, we
choose the path represented in blue in Fig.~\ref{fig:ACandGACE}, leading
thus to 
\begin{eqnarray}\label{eq:Exc_AC_wANDlambda}
{E}^w_{\rm xc}[n]&=&
\int^1_0\ddroit
\lambda\,\frac{\ddroit F^{\lambda,w}[n]}{\ddroit\lambda}
-E_{\rm H}[n]
\nonumber\\
&=&
\int^1_0\ddroit\lambda\,\left[
\frac{\ddroit F^{\lambda,0}[n]}{\ddroit\lambda}
+\int_0^w
\ddroit \xi\,
\frac{\ddroit^2
F^{\lambda,\xi}[n]}{ \ddroit \xi\ddroit\lambda}
\right]
-E_{\rm H}[n]
\nonumber\\
&=&
{E}_{\rm xc}[n]+
\int_0^w
\ddroit \xi\,
\left(
\frac{\ddroit
F^{1,\xi}[n]}{ \ddroit \xi}
-
\frac{\ddroit
F^{0,\xi}[n]}{ \ddroit \xi}
\right)
,
\end{eqnarray}
where the partially-interacting GOK functional equals along the GACE
\begin{eqnarray}\label{Fxlambda_def}
F^{\lambda,\xi}[n]&=&
(1-\xi)\,\langle\Psi^{\lambda,\xi}_1\vert\hat{T}+\lambda\hat{W}_{\rm
ee}\vert\Psi^{\lambda,\xi}_1\rangle
+\xi\,\langle\Psi^{\lambda,\xi}_2\vert\hat{T}+\lambda\hat{W}_{\rm
ee}\vert\Psi^{\lambda,\xi}_2\rangle.
\end{eqnarray}
Since, according to Appendix~\ref{appendix:deriv_Flambdax},
\begin{eqnarray}\label{dFlambax}
\displaystyle 
\frac{\ddroit F^{\lambda,\xi}[n]}{\ddroit
\xi}&=&\mathcal{E}^{\lambda,\xi}_2-\mathcal{E}^{\lambda,\xi}_1,
\end{eqnarray}
we finally obtain
\begin{eqnarray}\label{eq:Exc_AC_wANDlambda_2}
{E}^w_{\rm xc}[n]
&=&
{E}_{\rm xc}[n]+
\int_0^w \ddroit \xi\,
\Bigg[\Big(\mathcal{E}^{1,\xi}_2-\mathcal{E}^{1,\xi}_1\Big)
-
\Big(\mathcal{E}^{0,\xi}_2-\mathcal{E}^{0,\xi}_1\Big)
\Bigg]
.
\end{eqnarray}
The exact deviation of the ensemble exchange--correlation energy from the
ground-state one is therefore obtained by integrating the difference in excitation
energies 
\begin{eqnarray}\label{eq:diff_physical_KS_XE}
\Delta_{\rm xc}^{\xi}[n]=
\Big(\mathcal{E}^{1,\xi}_2-\mathcal{E}^{1,\xi}_1\Big)
-
\Big(\mathcal{E}^{0,\xi}_2-\mathcal{E}^{0,\xi}_1\Big)
\end{eqnarray}
between the physical and non-interacting GOK systems 
over the weight interval $[0,w]$ while keeping 
the ensemble density fixed. Equivalently, $\Delta_{\rm xc}^{\xi}[n]$ is
the first-order derivative of the ensemble exchange--correlation energy: 
\begin{eqnarray}\label{eq:1stderiv_diff_physical_KS_XE}
\Delta_{\rm xc}^{\xi}[n]=\frac{\ddroit{E}^{\xi}_{\rm xc}[n]}{ \ddroit \xi}.
\end{eqnarray}
According to Eqs.~(\ref{eq:Exc_AC_wANDlambda_2}) and (\ref{eq:diff_physical_KS_XE}), 
the ensemble exchange--correlation energy can be expanded through
second order in  
$w$ as follows,
\begin{eqnarray}\label{eq:Exc_AC_wANDlambda_Taylorexp}
{E}^w_{\rm xc}[n]
&=&
{E}_{\rm xc}[n]+w\Delta_{\rm
xc}^0[n]+\frac{w^2}{2}
\left.\frac{\ddroit \Delta_{\rm xc}^{\xi}[n]}{\ddroit
\xi}\right|_{\xi=0}
+\mathcal{O}(w^3)
,
\end{eqnarray}
where the first-order Taylor expansion coefficient 
\begin{eqnarray}\label{eq:taylorexp1storder}
\Delta_{\rm xc}^0[n]=
\Big(\mathcal{E}^{1,0}_2-\mathcal{E}^{1,0}_1\Big)
-
\Big(\mathcal{E}^{0,0}_2-\mathcal{E}^{0,0}_1\Big)
\end{eqnarray}
can be rewritten
more explicitly
as
\begin{eqnarray}\label{eq:taylorexp1storder_2}
\Delta_{\rm xc}^0[n]=
\Big(E_2[n]-E_1[n]\Big)
-
\Big(\varepsilon^0_2[n]-\varepsilon^0_1[n]\Big),
\end{eqnarray}
where, for convenience, the first excitation in the non-interacting KS system (to which
the GOK system reduces for $\xi=0$)  
is assumed to be a single excitation. The corresponding excitation
energy $\mathcal{E}^{0,0}_2-\mathcal{E}^{0,0}_1$ is then equal to the
HOMO-LUMO gap 
$\varepsilon^0_2[n]-\varepsilon^0_1[n]$ in the KS system whose ground-state density equals $n$. In case of
multiple excitations the excitation energy would simply be written as the
sum of KS orbital energy differences.
On the other hand, the first excitation energy
$\mathcal{E}^{1,0}_2-\mathcal{E}^{1,0}_1$ 
in the fully-interacting system whose ground-state density equals $n$
is simply denoted $E_2[n]-E_1[n]$.

Note that, in the particular case
where $n$ equals the exact ground-state density $n^0=n_{\tilde{\Psi}_1}$ of the true
physical system that is described by the Schr\"{o}dinger
Eq.~(\ref{Schroeqtruesyst}), the exact excitation energy $E_2-E_1$ and the
conventional KS HOMO-LUMO gap $\varepsilon^0_2-\varepsilon^0_1$ are
recovered, leading thus to    
\begin{eqnarray}\label{eq:taylorexp1storder_GSdensity}
\Delta_{\rm xc}^0[n^0]=
\Big(E_2-E_1\Big)
-
\Big(\varepsilon^0_2-\varepsilon^0_1\Big).
\end{eqnarray}
Levy~\cite{PRA_Levy_XE-N-N-1} has shown that the term on the left-hand side of
Eq.~(\ref{eq:taylorexp1storder_GSdensity}) can be interpreted as a
discontinuous change in the exchange--correlation potential as
$w\rightarrow0$. For that
reason we will refer to $\Delta_{\rm xc}^{\xi}[n]$ as the 
exchange--correlation {\it derivative
discontinuity} (DD) density functional in the following.    

Let us now focus on the second-order Taylor expansion coefficient in
Eq.~(\ref{eq:Exc_AC_wANDlambda_Taylorexp}). Since, according to
Eq.~(\ref{acw}) and the
Hellmann--Feynman theorem, 
\begin{eqnarray}\label{eq:HFthweight}
\frac{\ddroit\mathcal{E}_i^{\lambda,\xi}}{ \ddroit \xi}=
\int \ddroit{\bf r}\,\frac{\partial v^{\lambda,\xi}({\bf
r})}{\partial \xi}\,{n}_{\Psi^{\lambda,\xi}_i}({\bf r})
,\hspace{0.4cm}
i=1,2,
\end{eqnarray}
the first-order derivative of the exchange--correlation DD functional can be
expressed, according to Eq.~(\ref{eq:diff_physical_KS_XE}), as 
\begin{eqnarray}\label{eq:taylorexp2ndorder}
\left.\frac{\ddroit \Delta_{\rm xc}^{\xi}[n]}{\ddroit
\xi}\right|_{\xi=0}
&=&
\int \ddroit{\bf r}\,\left.\frac{\partial v^{1,\xi}({\bf
r})}{\partial \xi}\right|_{\xi=0}\,\Big({n}_{\Psi^{1,0}_2}({\bf r})-
{n}_{\Psi^{1,0}_1}({\bf r})
\Big)
\nonumber\\
&&-\int \ddroit{\bf r}\,\left.\frac{\partial v^{0,\xi}({\bf
r})}{\partial \xi}\right|_{\xi=0}\,\Big({n}_{\Psi^{0,0}_2}({\bf r})-
{n}_{\Psi^{0,0}_1}({\bf r})
\Big).
\end{eqnarray}
With the notations of Eqs.~(\ref{Fwdef}) and (\ref{Twsdef}), 
the first excited states of
the fully-interacting ($\Psi^{1,0}_2$) and KS ($\Psi^{0,0}_2$) systems whose
ground-state densities equal $n$ simply correspond to $\Psi^0_2[n]$ and $\Phi^0_2[n]$,
respectively. As
${n}_{\Psi^{1,0}_1}={n}_{\Psi^{0,0}_1}=n$, according to the density
constraint in Eq.~(\ref{ACwdensconstraints}), we obtain 
\begin{eqnarray}\label{eq:taylorexp2ndorder_2}
\left.\frac{\ddroit \Delta_{\rm xc}^{\xi}[n]}{\ddroit
\xi}\right|_{\xi=0}
&=&
\int \ddroit{\bf r}\,\left.\frac{\partial v^{1,\xi}({\bf
r})}{\partial \xi}\right|_{\xi=0}\,\Big({n}_{\Psi^0_2[n]}({\bf r})-
{n}({\bf r})
\Big)
\nonumber\\
&&-\int \ddroit{\bf r}\,\left.\frac{\partial v^{0,\xi}({\bf
r})}{\partial \xi}\right|_{\xi=0}\,\Big({n}_{\Phi^0_2[n]}({\bf r})-
{n}({\bf r})
\Big).
\end{eqnarray}
Moreover, as shown in Appendix~\ref{appendix:GOKpot}, the
fully-interacting and
GOK local potentials are
connected as follows
\begin{eqnarray}\label{eq:vreal_KS_connection}
v^{0,\xi}({\bf r})=v^{1,\xi}({\bf r})+
\frac{\delta {E}^{\xi}_{\rm Hxc}}{\delta n({\bf r})}[n]
,
\end{eqnarray}
which leads to the final expression
\begin{eqnarray}\label{eq:taylorexp2ndorder_3}
\left.\frac{\ddroit \Delta_{\rm xc}^{\xi}[n]}{\ddroit
\xi}\right|_{\xi=0}
&=&
\int \ddroit{\bf r}\,\left.\frac{\partial 
}{\partial \xi}
\frac{\delta {E}^{\xi}_{\rm xc}}{\delta n({\bf r})}[n]
\right|_{\xi=0}\,\Big(
{n}({\bf r})-{n}_{\Psi^0_2[n]}({\bf r})
\Big)
\nonumber\\
&&+\int \ddroit{\bf r}\,\left.\frac{\partial v^{0,\xi}({\bf
r})}{\partial \xi}\right|_{\xi=0}\,\Big({n}_{\Psi^0_2[n]}({\bf r})-{n}_{\Phi^0_2[n]}({\bf r})
\Big),
\end{eqnarray}
or, equivalently, according to Eq.~(\ref{eq:1stderiv_diff_physical_KS_XE}),
\begin{eqnarray}\label{eq:taylorexp2ndorder_4}
\left.\frac{\ddroit \Delta_{\rm xc}^{\xi}[n]}{\ddroit
\xi}\right|_{\xi=0}
&=&
\int \ddroit{\bf r}\,
\frac{\delta {\Delta}^0_{\rm xc}}{\delta n({\bf r})}[n]
\,\Big(
{n}({\bf r})-{n}_{\Psi^0_2[n]}({\bf r})
\Big)
\nonumber\\
&&+\int \ddroit{\bf r}\,\left.\frac{\partial v^{0,\xi}({\bf
r})}{\partial \xi}\right|_{\xi=0}\,\Big({n}_{\Psi^0_2[n]}({\bf r})-{n}_{\Phi^0_2[n]}({\bf r})
\Big).
\end{eqnarray}

Note that the Hartree density-functional potential does not appear in
the first term on the right-hand side of 
Eq.~(\ref{eq:taylorexp2ndorder_3}) since the ensemble Hartree
density-functional energy is weight-independent (see Eq.~(\ref{complementHensembledensity})). In addition, in
the particular case where $n$ equals the exact ground-state density
$n^0$ of the true physical system, Eq.~(\ref{eq:taylorexp2ndorder_4}) becomes
\begin{eqnarray}\label{eq:taylorexp2ndorder_GSdensity_2}
\left.\frac{\ddroit \Delta_{\rm xc}^{\xi}[n^0]}{\ddroit
\xi}\right|_{\xi=0}
&=&
\int \ddroit{\bf r}\,
\frac{\delta \Delta^0_{\rm xc}}{\delta n({\bf r})}[n^0]
\,\Big(
{n^0}({\bf r})-{n}_{\tilde{\Psi}_2}({\bf r})
\Big)
\nonumber\\
&&+
\int \ddroit{\bf r}\,\left.\frac{\partial v_{\rm s}^{\xi}[n^0]({\bf
r})}{\partial \xi}\right|_{\xi=0}\,\Big({n}_{\tilde{\Psi}_2}({\bf
r})-{n}_{\tilde{\Phi}^0_2}({\bf r})
\Big),
\end{eqnarray}
where $\tilde{\Psi}_2$ and $\tilde{\Phi}^0_2$ are the first excited
states of the physical and KS systems, respectively (see
Eqs.~(\ref{Schroeqtruesyst})
and (\ref{sceqensembleKS})). 
For clarity the local GOK
potential for which the ensemble density remains equal to the exact ground-state
density $n^0$ as the weight of
the ensemble varies in the vicinity of $\xi=0$ has been denoted $v_{\rm s}^{\xi}[n^0]$.
As shown in the next section, the Taylor expansion we obtained within
the GACE for the ensemble
exchange--correlation energy leads to
stringent constraints on the functional. 

\subsection{Exact ensemble and excitation
energies}\label{subsec:stringent_constraints}

Let us consider the GACE in the particular case where the density $n$ equals the exact ensemble density
$n^w$ of the physical system. According to Eqs.~(\ref{Schroeqtruesyst}) and
(\ref{sceqensembleKS}), the local potentials $v^{1,w}({\bf r})$ and
$v^{0,w}({\bf r})$ correspond then to the nuclear
$v_{\rm ne}({\bf r})$ and GOK
$v_{\rm ne}({\bf r})+\delta {E}^w_{\rm
Hxc}[n^w]/\delta n({\bf r})$ 
potentials, respectively. Consequently, the fully-interacting excitation energy becomes
the true physical one $E_2-E_1$, while the non-interacting excitation
energy
is the GOK one obtained from
Eq.~(\ref{sceqensembleKS}), leading thus to the following expression for
the ensemble exchange--correlation DD energy: 
\begin{eqnarray}\label{eq:XE_Xe_DD}
\Delta_{\rm
xc}^w[n^w]=E_2-E_1-\Big(
\mathcal{E}^w_{{\rm s},2}
-\mathcal{E}^w_{{\rm s},1}
\Big).
\end{eqnarray}
When the first excitation in the GOK system corresponds to a single
excitation, the corresponding excitation energy can be rewritten as an
orbital energy difference  
\begin{eqnarray}\label{eq:KSXE_orb_energies}
\mathcal{E}^w_{{\rm s},2}
-\mathcal{E}^w_{{\rm s},1}
=
\varepsilon_{2}^w-\varepsilon_{1}^w,
\end{eqnarray}
and, consequently, the expression of Gross \etal~\cite{PRA_GOK_EKSDFT} for the exact first
excitation energy is recovered:
\begin{eqnarray}\label{XEexact}
\displaystyle 
E_2-E_1&=&
\varepsilon_{2}^w-\varepsilon_{1}^w+\Delta_{\rm xc}^w[n^w]
\nonumber\\
&=&
\left.
\varepsilon_{2}^w-\varepsilon_{1}^w
+\frac{\ddroit E_{\rm
xc}^{\xi}[n^w]}{\ddroit
\xi}\right|_{\xi=w}.
\end{eqnarray}
It becomes clear from Eq.~(\ref{XEexact}) that the weight-dependent
exchange--correlation DD density functional $\Delta_{\rm xc}^w[n]$
plays a crucial role in the calculation of excitation energies in
GOK-DFT. 

In the rest of this work we will show how the GACE could be used for 
the development of ensemble DFAs. 
Before, let us mention that stringent constraints on the density
functional $\Delta_{\rm xc}^w[n]$ can be derived from
Eq.~(\ref{XEexact}) when rewriting, according to Eq.~(\ref{ensembleener}), the    
excitation energy as the first-order derivative of the ensemble
energy with respect to the ensemble weight $w$:
\begin{eqnarray}\label{eq:dEdw=XE}
\frac{\ddroit {E}^w}{\ddroit w}=E_2-E_1, \hspace{0.6cm} 0\leq w \leq
\frac{1}{2}.
\end{eqnarray}
In the exact theory this derivative 
should therefore not vary with $w$ or, equivalently, the ensemble energy
should have no curvature:
\begin{eqnarray}\label{eq:Enocurvature}
\frac{\ddroit^k {E}^w}{\ddroit w^k}=0, \hspace{0.6cm} 0\leq w \leq
\frac{1}{2},\hspace{0.6cm} k\geq 2.
\end{eqnarray}
Note that differentiability with respect to the ensemble weight $w$ will
be
assumed (but it is in principle not guaranteed) for individual terms on the right-hand side of
Eq.~(\ref{XEexact}).

For the purpose of constructing ensemble DFAs from regular
ground-state DFAs, as proposed by Nagy~\cite{Nagy_enseXpot} and
Paragi~\etal~\cite{ParagiXens,ParagiXCens}, Eqs.~(\ref{eq:dEdw=XE}) and (\ref{eq:Enocurvature})
should be taken in the $w=0$ limit. Here we will consider derivatives through second order only
($k=2$), which leads to the two exact
conditions
\begin{eqnarray}\label{eq:twocondensE_w=0}
\left.\frac{\ddroit {E}^w}{\ddroit w}\right|_{w=0}=E_2-E_1, 
\end{eqnarray}
and 
\begin{eqnarray}\label{eq:twocondensE_w=0_2}
\left.\frac{\ddroit^2 {E}^w}{\ddroit w^2}\right|_{w=0}=0. 
\end{eqnarray}
Since, according to Eq.~(\ref{ensembledens}), the ensemble
exchange--correlation DD energy is expanded through first order as
\begin{eqnarray}\label{Deltaxc_talorexpw}
\displaystyle 
\Delta_{\rm xc}^w[n^w]
&=&
\Delta_{\rm xc}^w[n^0]
+
\int \ddroit{\bf r}\,\frac{\delta \Delta_{\rm xc}^w}{\delta n({\bf
r})}[n^0]\Big(n^w({\mathbf r})-n^0({\mathbf r})\Big)
+\mathcal{O}(w^2)
\nonumber
\\
&=&
\Delta_{\rm xc}^0[n^0]+w
\Bigg(\left.\frac{\ddroit \Delta_{\rm xc}^{\xi}[n^0]}{\ddroit
\xi}\right|_{\xi=0}
+
\int \ddroit{\bf r}\,\frac{\delta \Delta_{\rm xc}^0}{\delta n({\bf
r})}[n^0]\Big(n_{\tilde{\Psi}_2}({\mathbf r})-n^0({\mathbf r})\Big)\Bigg)
\nonumber
\\
&&+\mathcal{O}(w^2),
\end{eqnarray}
which gives, according to Eq.~(\ref{eq:taylorexp2ndorder_GSdensity_2}),
\begin{eqnarray}\label{Deltaxc_talorexpw_2}
\displaystyle 
\Delta_{\rm xc}^w[n^w]
&=&
\Delta_{\rm xc}^0[n^0]+w
\int \ddroit{\bf r}\,\left.\frac{\partial v_{\rm s}^{\xi}[n^0]({\bf
r})}{\partial \xi}\right|_{\xi=0}\,\Big({n}_{\tilde{\Psi}_2}({\bf
r})-{n}_{\tilde{\Phi}^0_2}({\bf r})\Big)
+\mathcal{O}(w^2),
\end{eqnarray}
we obtain through first order, from Eqs.~(\ref{XEexact}) and (\ref{eq:dEdw=XE}), 
\begin{eqnarray}\label{XEexact_talorexpw_2}
\displaystyle 
&&\frac{\ddroit {E}^w}{\ddroit w}
=
\Big(\varepsilon^0_{2}-\varepsilon^0_{1}\Big)+
\Delta_{\rm xc}^0[n^0]
\nonumber\\
&&
+w\Bigg[\left.
\frac{\ddroit }{\ddroit w}
\Big(\varepsilon_{2}^w-\varepsilon_{1}^w\Big)
\right|_{w=0}
+
\int \ddroit{\bf r}\,\left.\frac{\partial v_{\rm s}^{\xi}[n^0]({\bf
r})}{\partial \xi}\right|_{\xi=0}\,\Big({n}_{\tilde{\Psi}_2}({\bf
r})-{n}_{\tilde{\Phi}^0_2}({\bf r})\Big)\Bigg]
\nonumber\\
&&
+\mathcal{O}(w^2).
\end{eqnarray}
Eq.~(\ref{eq:taylorexp1storder_GSdensity}) is thus recovered from 
Eq.~(\ref{eq:twocondensE_w=0}) while the second constraint
in Eq.~(\ref{eq:twocondensE_w=0_2}) leads to  
\begin{eqnarray}\label{eq:2ndexactcondition}
\left.
\frac{\ddroit }{\ddroit w}
\Big(\varepsilon_{2}^w-\varepsilon_{1}^w\Big)
\right|_{w=0}
=-
\int \ddroit{\bf r}\,\left.\frac{\partial v_{\rm s}^{\xi}[n^0]({\bf
r})}{\partial \xi}\right|_{\xi=0}\,\Big({n}_{\tilde{\Psi}_2}({\bf
r})-{n}_{\tilde{\Phi}^0_2}({\bf r})\Big)
.
\end{eqnarray}
By rewriting the derivative on the left-hand side of
Eq.~(\ref{eq:2ndexactcondition}),
according to the Hellmann--Feynman theorem and
Eqs.~(\ref{sceqensembleKS}) and (\ref{eq:KSXE_orb_energies}), as
\begin{eqnarray}\label{eq:ddeltaedw0}
\left.
\frac{\ddroit }{\ddroit w}
\Big(\varepsilon_{2}^w-\varepsilon_{1}^w\Big)
\right|_{w=0}
&=&
\left\langle \tilde{\Phi}_2^0\left|
\left.
\frac{\ddroit
}{\ddroit w}
 \hat{{V}}^w_{\rm Hxc}[n^w]
\right
|
_{w=0}
\right|\tilde{\Phi}_2^0\right\rangle
\nonumber\\
&&
-
\left\langle \tilde{\Phi}_1^0\left|
\left.
\frac{\ddroit
}{\ddroit w}
 \hat{{V}}^w_{\rm Hxc}[n^w]
\right
|
_{w=0}
\right|\tilde{\Phi}_1^0\right\rangle
\nonumber\\
&=&
\int {\rm d}{\mathbf r}\,
\left.
\frac{\ddroit}{\ddroit w}
\frac{\delta E_{\rm
Hxc}^w}{\delta n({\mathbf r})}[n^w]
\right
|
_{w=0}
\Big(n_{\tilde{\Phi}^0_2}({\mathbf r})-n^0({\mathbf
r})\Big),
\end{eqnarray}
and using
\begin{eqnarray}\label{eq:simplify_dvxnx_dx}
\left.\frac{\ddroit}{\ddroit w}
\frac{\delta E_{\rm
Hxc}^w}{\delta n({\mathbf r})}[n^w]
\right|_{w=0}&=&
\left.\frac{\partial 
}{\partial w}
\frac{\delta {E}^w_{\rm xc}}{\delta n({\bf r})}[n^0]
\right|_{w=0}
\nonumber\\
&&+
\int
\ddroit{\bf r'}\,
K_{\rm Hxc}({\bf r'},{\bf r})
\Big(n_{\tilde{\Psi}_2}({\mathbf r'})-n^0({\mathbf r'})\Big),
\end{eqnarray}
where $K_{\rm Hxc}({\bf r'},{\bf
r})=\delta^2E_{\rm Hxc}[n^0]/\delta n({\mathbf r'})\delta n({\mathbf r})$
denotes the ground-state Hxc kernel, 
we conclude from Eq.~(\ref{eq:2ndexactcondition}) that the exact
constraint in Eq.~(\ref{eq:twocondensE_w=0_2}) is equivalent to 
\begin{eqnarray}\label{eq:ddeltaedw0_2}
&&\int {\rm d}{\mathbf r}\,
\frac{\delta \Delta^0_{\rm
xc}}{\delta n({\mathbf r})}[n^0]
\Big(n_{\tilde{\Phi}^0_2}({\mathbf r})-n^0({\mathbf
r})\Big)
\nonumber\\
&&+
\int
\int 
\ddroit{\bf r}
\ddroit{\bf r'}\,
K_{\rm Hxc}({\bf r'},{\bf r})
\Big(n_{\tilde{\Psi}_2}({\mathbf r'})-n^0({\mathbf r'})\Big)
\Big(n_{\tilde{\Phi}^0_2}({\mathbf r})-n^0({\mathbf
r})\Big)
\nonumber\\
&&=
-
\int \ddroit{\bf r}\,\left.\frac{\partial v_{\rm s}^{\xi}[n^0]({\bf
r})}{\partial \xi}\right|_{\xi=0}\,\Big({n}_{\tilde{\Psi}_2}({\bf
r})-{n}_{\tilde{\Phi}^0_2}({\bf r})\Big).
\end{eqnarray}
Note that, when simplifying the Hartree contribution only 
in Eq.~(\ref{eq:simplify_dvxnx_dx}),
relation~(\ref{eq:ddeltaedw0_2}) can alternatively be rewritten as 
\begin{eqnarray}\label{eq:ddeltaedw0_Levy_form}
&&
-\int {\rm d}{\mathbf r}\,
\left.
\frac{\ddroit}{\ddroit w}
\frac{\delta E_{\rm
xc}^w}{\delta n({\mathbf r})}[n^w]
\right
|
_{w=0}
\Big(n_{\tilde{\Phi}^0_2}({\mathbf r})-n^0({\mathbf
r})\Big)
\nonumber\\
&&
-
\int \ddroit{\bf r}\,\left.\frac{\partial v_{\rm s}^{\xi}[n^0]({\bf
r})}{\partial \xi}\right|_{\xi=0}\,\Big({n}_{\tilde{\Psi}_2}({\bf
r})-{n}_{\tilde{\Phi}^0_2}({\bf r})\Big)
\nonumber\\
&&=
\int
\ddroit{\bf r}\,
\left(\Big[n_{\tilde{\Phi}^0_2}({\mathbf r})-n^0({\mathbf
r})\Big]
\int 
\ddroit{\bf r'}\,
\frac{\Big[n_{\tilde{\Psi}_2}({\mathbf r'})-n^0({\mathbf r'})\Big]}
{\vert{\mathbf r}-{\mathbf r'}\vert}
\right),
\end{eqnarray}
which is nothing but Levy's constraint (see Eq.~(30) in
Ref.~\cite{PRA_Levy_XE-N-N-1}) in the
$w\rightarrow 0$ limit. Interestingly, we obtain in
the second integral on the left-hand side of
Eq.~(\ref{eq:ddeltaedw0_Levy_form}) an
explicit expression for the contribution that
arises from the
discontinuous change of the exchange--correlation potential as 
$w\rightarrow 0$. Note that this contribution comes directly from  
the GACE, where the ensemble density of the non-interacting system is held fixed to the ground-state
density $n^0$ while the ensemble weight varies in the vicinity of $\xi=0$.

Returning to the formulation in Eq.~(\ref{eq:ddeltaedw0_2}), an accurate
value for the integral on the right-hand side could in principle be
obtained 
when constructing the GACE with {\it ab initio} methods, as discussed
further in
Sec.~\ref{subsec:construct_gace}. The contributions on the left-hand
side of Eq.~(\ref{eq:ddeltaedw0_2}) can, on the other hand, be
computed with DFAs.
The
stringent constraint we derived could thus be used for developing DFAs to
$\Delta^0_{\rm xc}[n]$. Interestingly the ground-state kernel, that
plays a key role in TD-DFT~\cite{Casida_tddft_review_2012}, appears in the derivation of the
excitation energy within GOK-DFT. Connections between the two approaches
should be investigated further in the light 
of the recent work of Ziegler and
coworkers~\cite{JCP09_Ziegler_relation_TD-DFT_VDFT,JCTC13_Ziegler_SCF-CV-DFT} on {\it constricted
variational density-functional theory} (CV-DFT). A formal connection
might also be obtained when considering imaginary temperatures in
Boltzmann factors for the ensemble weights~\cite{PRA13_Pernal_srEDFT}. Work is in progress
in these directions.

\subsection{Construction of the 
GACE
}\label{subsec:construct_gace}

By analogy with traditional ground-state
AC calculations~\cite{Teale:2009p2020,Teale:2010,Teale:2010b}, the GACE
could in principle be
constructed from the partially-interacting GOK functional
introduced in Eq.~(\ref{Fxlambda_def}). Note that the functional is defined for ensemble $v$-representable
densities. The domain of the functional can be enlarged to ensemble $N$
representable densities by using a Levy--Lieb 
constrained-search formulation~\cite{PRA_GOK_EKSDFT,PRA13_Pernal_srEDFT},
\begin{eqnarray}\label{Flambda_x_LL}
F^{\lambda,\xi}[n]&=&\underset{\{\Psi_1,\Psi_2\}^{\xi}\rightarrow n}{\rm
min}\bigg\{
(1-\xi)\,\langle\Psi_1\vert\hat{T}+\lambda\hat{W}_{\rm ee}\vert\Psi_1\rangle
+\xi\,\langle\Psi_2\vert\hat{T}+\lambda\hat{W}_{\rm
ee}\vert\Psi_2\rangle\bigg\},
\end{eqnarray}
where the minimization in Eq.~(\ref{Flambda_x_LL}) is restricted to orthonormal sets
of wavefunctions $\{\Psi_1,\Psi_2\}^{\xi}$ whose ensemble density
$(1-\xi)\,n_{\Psi_1}+\xi\,n_{\Psi_2}$ equals $n$. 

The minimizing
wavefunctions $\Psi^{\lambda,\xi}_1$ and $\Psi^{\lambda,\xi}_2$ can
alternatively be reached when searching for the local potential
$v^{\lambda,\xi}$ that was introduced in Eq.~(\ref{acw}). For that purpose
we define, for a
given local potential $v$, the partially-interacting Hamiltonian
$\hat{H}^\lambda[v]=\hat{T}+\lambda\hat{W}_{\rm ee}+
\int {\rm d}{\bf
r}\,v({\bf r})\,\hat{n}({\bf r})
$ and denote $\mathcal{E}^\lambda_1[v]$ and
$\mathcal{E}^\lambda_2[v]$ the associated ground- and first-excited-state
energies, respectively. According to the GOK variational
principle,
\begin{eqnarray}\label{GOK_principle_lambda_x}
(1-\xi)\,\langle\Psi^{\lambda,\xi}_1\vert
\hat{H}^\lambda[v]\vert\Psi^{\lambda,\xi}_1\rangle
+\xi\,\langle\Psi^{\lambda,\xi}_2\vert
\hat{H}^\lambda[v]\vert\Psi^{\lambda,\xi}_2\rangle\geq
(1-\xi)\mathcal{E}^\lambda_1[v]+\xi\,\mathcal{E}^\lambda_2[v],
\end{eqnarray}
or, equivalently,
\begin{eqnarray}\label{GOK_principle_lambda_x_2}
F^{\lambda,\xi}[n]\geq
(1-\xi)\mathcal{E}^\lambda_1[v]+\xi\,\mathcal{E}^\lambda_2[v]
-
\int {\rm d}{\bf
r}\,v({\bf r})\,{n}({\bf r})
.
\end{eqnarray}
The partially-interacting GOK functional can therefore be
rewritten as a Legendre--Fenchel
transform~\cite{Eschrig,Kutzelnigg:2006,vanLeeuwen:2003,IJQC83_Lieb_LF_transf} 
\begin{eqnarray}\label{Legrendre_Fenchel_Flambda_x}
F^{\lambda,\xi}[n]=\underset{v}{\rm sup}
\bigg\{
\mathcal{F}^{\lambda,\xi}[v,n]
\bigg\}
,
\end{eqnarray}
where 
\begin{eqnarray}\label{Legrendre_Fenchel_Flambda_x_2}
\mathcal{F}^{\lambda,\xi}[v,n]
=
(1-\xi)\mathcal{E}^\lambda_1[v]+\xi\,\mathcal{E}^\lambda_2[v]
-
\int {\rm d}{\bf
r}\,v({\bf r})\,{n}({\bf r}),
\end{eqnarray}
and the maximizing potential in Eq.~(\ref{Legrendre_Fenchel_Flambda_x}),
if it exists, equals $v^{\lambda,\xi}$. In the latter case, where we
assume that the density $n$ can be represented by a non-degenerate
two-state partially-interacting ensemble, expressions
in Eqs.~(\ref{Flambda_x_LL}) and (\ref{Legrendre_Fenchel_Flambda_x}) are
equivalent. In the special case $\xi=0$, it would therefore be assumed
that the density $n$ is pure-state-$v$-representable. For any density,
the ground-state Legendre--Fenchel transform
recovered when $\xi=0$ is in fact equivalent to the Levy--Valone--Lieb
functional~\cite{valone:4653}. Degeneracies associated with the ground-state energy
$\mathcal{E}^\lambda_1[v]$ can indeed allow for the description of densities that
are not pure-state-$v$-representable.

Returning to non-degenerate two-state ensemble $v$-representable
densities, we note that Nagy's AC~\cite{Nagy_ensAC} can be constructed by fixing the
ensemble weight $\xi$ to a
given value $w$ and by choosing the
weight-dependent ensemble density $n^w$ as input
density in Eq.~(\ref{Legrendre_Fenchel_Flambda_x}). In this case the maximizing local
potential $\tilde{v}^{\lambda,w}$ is determined from the stationary 
condition 
\begin{eqnarray}\label{Legrendre_Fenchel_Flambda_w_varcond}
\frac{\delta \mathcal{F}^{\lambda,w}}{\delta v(\bf r)}
 [\tilde{v}^{\lambda,w},n^w]=0.
\end{eqnarray}
On the other hand, the GACE is constructed when varying both
ensemble weight and interaction strength while keeping the density fixed
to $n$. The maximizing potential $v^{\lambda,\xi}$ is then obtained from the variational
condition
\begin{eqnarray}\label{Legrendre_Fenchel_Flambda_x_varcond}
\frac{\delta \mathcal{F}^{\lambda,\xi}}{\delta v(\bf r)}
 [{v}^{\lambda,\xi},n]=0,
 \hspace{0.4cm} 0\leq \xi\leq w,
\end{eqnarray}
which, according to Eq.~(\ref{Legrendre_Fenchel_Flambda_x_2}), is equivalent to  
\begin{eqnarray}\label{Legrendre_Fenchel_Flambda_x_varcond_2}
(1-\xi)
\frac{\delta \mathcal{E}^\lambda_1}{\delta {v}({\bf
r})}[{v}^{\lambda,\xi}]
+\xi\,
\frac{\delta \mathcal{E}^\lambda_2}{\delta {v}({\bf
r})}[{v}^{\lambda,\xi}]
={n}({\bf r}),
 \hspace{0.4cm} 0\leq \xi\leq w.
\end{eqnarray}
Since, according to the Hellmann--Feynman theorem, each individual functional derivatives correspond to the individual
densities,
\begin{eqnarray}\label{individual_densities}
\frac{\delta \mathcal{E}^\lambda_i}{\delta {v}({\bf
r})}[{v}^{\lambda,\xi}]
={n}_{\Psi_i^{\lambda,\xi}}({\bf r}),
 \hspace{0.4cm} \hspace{0.2cm} i=1,2,
\end{eqnarray}
the density constraint in Eq.~(\ref{ACwdensconstraints}) is recovered
from Eq.~(\ref{Legrendre_Fenchel_Flambda_x_varcond_2}).

Let us consider the particular case where the input density $n$ equals
$n^w$. 
In contrast to Nagy's AC, the variational condition in
Eq.~(\ref{Legrendre_Fenchel_Flambda_x_varcond}) will be fulfilled
along the GACE {\it for
any value of the ensemble weight} in the range $0\leq \xi\leq w$,
\begin{eqnarray}\label{Legrendre_Fenchel_Flambda_x_varcond_nw}
\frac{\delta \mathcal{F}^{\lambda,\xi}}{\delta v(\bf r)}
 [{v}^{\lambda,\xi},n^w]=0.
\end{eqnarray}
Nagy's AC is simply recovered when $\xi=w$. In this case 
${v}^{\lambda,\xi}$ reduces to the local potential  
$\tilde{v}^{\lambda,w}$ introduced in
Eq.~(\ref{Legrendre_Fenchel_Flambda_w_varcond}).

The GACE could in principle be computed along those lines by using {\it
ab initio} methods for the description of the partially-interacting
ensemble. For that purpose, the recent work of
Teale~\etal~\cite{Teale:2009p2020,Teale:2010,Teale:2010b} on the computation of ground-state ACs should be extended
to ensembles. Such an approach would provide precious data for the
development of ensemble DFAs.

Let us finally stress that the GACE offers some flexibility
in the choice of the input density. For convenience, one may wish to   
construct a GACE where the local potential ${v}^{\lambda,\xi}$ does not depend
on the ensemble weight $\xi$. Consequently, individual densities of the
ground- and first-excited states in the partially-interacting system would
be weight-independent. Since the ensemble density is fixed along the GACE, it
would simply mean that the individual  
densities are equal. As an illustration, we propose in the
following to construct such a GACE
analytically for the simple H$_2$ model system in a minimal basis.

\section{Analytical derivation of the GACE for H$_2$ in a minimal
basis}\label{subsec:theory:AC_min_basis}

We consider in this section the H$_2$ molecule in a Slater minimal basis
consisting of the $1s_A$ and $1s_B$ atomic orbitals localized on the
left and right hydrogen
atoms, respectively~\cite{H2minbasissetJChemEduc,Sharkas_JCP12_MCH}. The basis functions are identical with $\zeta=1$. For large bond distances the bonding and anti-bonding molecular
orbitals are equal to 
$1\sigma_g=\frac{1}{\sqrt{2}}\big(1s_A+1s_B\big)$
and
$1\sigma_u=\frac{1}{\sqrt{2}}\big(1s_A-1s_B\big)$, respectively. Both
traditional AC
and GACE will be constructed in the following within the $^1\Sigma_g^+$
symmetry. The space of two-electron wavefunctions to be considered
reduces then to the two Slater determinants $1\sigma_g^2$ and
$1\sigma_u^2$. Since these two determinants differ by a double excitation,
they are not
coupled by one-electron operators such as local potential 
operators. Even though equations are derived explicitly for H$_2$,
any two-level system that fulfils the latter 
condition could be described similarly. Returning to H$_2$,
in the dissociation limit, the two-state ensemble will
therefore consists of the neutral
$\frac{1}{\sqrt{2}}\big(1\sigma_g^2-1\sigma_u^2\big)$ and
ionic $\frac{1}{\sqrt{2}}\big(1\sigma_g^2+1\sigma_u^2\big)$ states.  

The analytical derivation of the Legendre--Fenchel transform is first presented
for the ground state in
Sec.~\ref{subsubsec:theory:AC_GS_min_basis}. The extension to the two-state
ensemble is then given in
Sec.~\ref{subsubsec:theory:AC_ensemble_min_basis}. In the light of these
derivations we finally propose in
Sec.~\ref{subsec:GSxcapprox} a simple DFA to the ensemble exchange--correlation
functional.

\subsection{AC for the ground state}\label{subsubsec:theory:AC_GS_min_basis}

Let the matrix representation of the physical fully-interacting Hamiltonian
in the basis of the $1\sigma_g^2$ and $1\sigma_u^2$ determinants be
\begin{eqnarray}\label{Hmatrix}
\left[ \hat{H} \right] =
\left[
\begin{array}{c c}
E_g & K \\
K & E_u
\end{array}
\right],
\end{eqnarray}
where 
\begin{eqnarray}\label{def_matrix_elements}
&&  E_i = \langle 1\sigma_i^2 \vert \hat{T} +\hat{W}_{\rm ee} +
  \hat{V}_{\rm ne} \vert 1\sigma_i^2 \rangle, \hspace{0.4cm} i=g,u,
  \nonumber\\
&& K = \langle 1\sigma_g^2 \vert \hat{W}_{\rm ee} \vert  1\sigma_u^2
\rangle.
\end{eqnarray}
The ground-state wavefunction $\Psi_1$ and ground-state energy $E_1$ are obtained by
diagonalizing $\left[ \hat{H} \right]$, which leads to 
\begin{eqnarray}\label{GSenergy}
E_1= \frac{1}{2} \left( {E}_g + {E}_u - \sqrt{\big({E}_g-{E}_u\big)^2 + 4 K^2} \right)
,
\end{eqnarray}
and
\begin{eqnarray}\label{GSwf}
\vert{\Psi}_1\rangle=\frac{1}{\sqrt{1+C_u^2}}\Big(\vert1\sigma_g^2\rangle+C_u\vert1\sigma_u^2\rangle\Big) 
,
\end{eqnarray}
with 
\begin{eqnarray}\label{GSCu}
C_u=\frac{E_1-E_g}{K}.
\end{eqnarray}
Since $1\sigma_g^2$ and $1\sigma_u^2$ differ by a double excitation,
they are not coupled by the density operator. Hence the ground-state density can be expressed as
\begin{eqnarray}\label{GSdens}
n^0({\bf r})=\langle{\Psi}_1 \vert\hat{n}({\bf r})\vert{\Psi}_1\rangle
=
\frac{1}{1+C_u^2}\Big(n_g({\bf r})+C_u^2\,n_u({\bf r})\Big)
,
\end{eqnarray}
where $n_g$ and $n_u$ denote the densities associated with the $1\sigma_g^2$
and $1\sigma_u^2$ determinants, respectively.

Before constructing the
AC for the ground-state density $n^0$, let us first mention that the
HK theorem may not be fulfilled in a finite basis~\cite{PRA83_Harriman_DFT_finite_basis,JMS10_Andreas_DFT_in_finite_basis}. Here a
non-interacting Hamiltonian $\hat{H}^0=\hat{T}+
\int {\rm d}{\bf
r}\,v({\bf r})\,\hat{n}({\bf r})
$ will simply be
represented by a diagonal  
matrix since the local potential operator does not couple
the $1\sigma_g^2$ and $1\sigma_u^2$ determinants: 
\begin{eqnarray}\label{HKSmatrix}
\left[ \hat{H}^0 \right] =
\left[
\begin{array}{c c}
\langle 1\sigma_g^2 \vert \hat{T}\vert1\sigma_g^2\rangle + V_g& 0 \\
0 & \langle 1\sigma_u^2 \vert \hat{T}\vert1\sigma_u^2\rangle + V_u 
\end{array}
\right],
\end{eqnarray}
where the two matrix elements 
$
V_g
$
and
$V_u$
defined as
\begin{eqnarray}\label{VgVudef}
V_i=
\int {\rm d}{\bf
r}\,v({\bf r})\,{n}_{i}({\bf r})
  , \hspace{0.4cm} i=g,u,
\end{eqnarray}
fully determine the potential in the minimal basis. In the particular
case where the density $n_g$ is considered, the KS local potential is obviously not
unique since the ground state of the non-interacting system remains equal
to the $1\sigma_g^2$ determinant as long as the following condition is
fulfilled 
\begin{eqnarray}\label{vks_notunique}
 V_g
- V_u 
<
\langle 1\sigma_u^2 \vert \hat{T}\vert1\sigma_u^2\rangle 
-\langle 1\sigma_g^2 \vert \hat{T}\vert1\sigma_g^2\rangle. 
\end{eqnarray}
On the other hand, the ground-state density $n^0$ is a linear
combination of $n_g$ and $n_u$. Consequently, the KS $1\sigma_g^2$ and
$1\sigma_u^2$ determinants must be degenerate so
that the non-interacting density equals the interacting one. In other
words an ensemble is required in the minimal basis while, in  
larger basis sets and for a finite bond distance, it is not (see, for
example, Ref.~\cite{Teale:2009p2020}). This
will be discussed further in the following. The 
KS potential is therefore uniquely defined (up to a constant) in the
minimal basis by the equality 
\begin{eqnarray}\label{vks_n0_unique}
 V^0_g
- V^0_u 
=
\langle 1\sigma_u^2 \vert \hat{T}\vert1\sigma_u^2\rangle 
-\langle 1\sigma_g^2 \vert \hat{T}\vert1\sigma_g^2\rangle. 
\end{eqnarray}

It is then relevant to construct an AC for the ground-state density
within the minimal basis. For
that purpose we introduce the matrix representation of the
partially-interacting Hamiltonian  
\begin{eqnarray}\label{Hlambdamatrix}
\left[ \hat{H}^\lambda \right] =
\left[
\begin{array}{c c}
\langle 1\sigma_g^2 \vert \hat{T}+
\lambda\hat{W}_{\rm
ee}
\vert1\sigma_g^2\rangle 
+ V_g& \lambda K \\
\lambda K & \langle 1\sigma_u^2 \vert \hat{T}
+
\lambda\hat{W}_{\rm
ee}
\vert1\sigma_u^2\rangle + V_u 
\end{array}
\right],
\end{eqnarray}
and, for convenience, substitute the parameters $\mathcal{V}_g$ and
$\mathcal{V}_u$ for $V_g$ and $V_u$, respectively, where 
\begin{eqnarray}\label{new_variables}
{V}_i=\lambda E_i -
\langle 1\sigma_i^2 \vert \hat{T}+
\lambda\hat{W}_{\rm
ee}
\vert1\sigma_i^2\rangle +\lambda\mathcal{V}_i, \hspace{0.4cm} i=g,u.
\end{eqnarray}
Note that one single
parameter
\begin{eqnarray}\label{def_upsilon}
\upsilon=\mathcal{V}_g-\mathcal{V}_u 
\end{eqnarray}
is in fact sufficient, since the local potential is determined up to a constant.
This leads to the following parameterization of the
partially-interacting Hamiltonian
\begin{eqnarray}\label{Hlambdamatrix_2}
\left[ \hat{H}^\lambda \right] =\lambda
\left[
\begin{array}{c c}
E_g+\upsilon& K \\
K & 
E_u
\end{array}
\right]+\lambda\mathcal{V}_u.
\end{eqnarray}
Note that, within this parameterization, the degeneracy of the KS
determinants is ensured for $\lambda=0$. 

The ground-state Legendre--Fenchel transform, from which we will construct
the AC for the ground-state density $n^0$, is obtained 
as follows~\cite{Eschrig,Kutzelnigg:2006,vanLeeuwen:2003}
\begin{eqnarray}\label{Legrendre_Fenchel_H2_Flambda_GS}
F^{\lambda}[n^0]=\underset{\upsilon}{\rm sup}
\bigg\{
\mathcal{F}^{\lambda}[\upsilon,n^0]
\bigg\}
,
\end{eqnarray}
where, according to Eqs.~(\ref{GSdens}) and (\ref{VgVudef}), 
\begin{eqnarray}\label{Legrendre_Fenchel_H2_Flambda_GS_def}
\mathcal{F}^{\lambda}[\upsilon,n^0]
&=&
\mathcal{E}^\lambda_1(\upsilon)-
\int {\rm d}{\bf
r}\,v({\bf r})\,{n}^0({\bf r}),
\nonumber
\\
&=&\mathcal{E}^\lambda_1(\upsilon)-
\frac{1}{1+C_u^2}\Big(V_g+C_u^2V_u\Big),
\end{eqnarray}
and, according to Eq.~(\ref{Hlambdamatrix_2}),
the auxiliary ground-state
energy equals
\begin{eqnarray}\label{GSauxienergy}
\mathcal{E}^\lambda_1(\upsilon)= \frac{\lambda}{2} \left( {E}_g +\upsilon+ {E}_u
- \sqrt{\big({E}_g+\upsilon-{E}_u\big)^2 + 4 K^2}
\right)+\lambda\mathcal{V}_u
.
\end{eqnarray}
Since in our parameterization $\mathcal{V}_u$ is a constant, $V_u$ does not vary with
$\upsilon$ and
\begin{eqnarray}\label{derivative_Vg}
\frac{\ddroit V_g}{\ddroit \upsilon}=\lambda,
\end{eqnarray}
according to Eqs.~(\ref{new_variables}) and (\ref{def_upsilon}).
The maximizing parameter $\upsilon^\lambda$ in
Eq.~(\ref{Legrendre_Fenchel_H2_Flambda_GS}) is therefore obtained
when solving
\begin{eqnarray}\label{MaxLegrendre_Fenchel_H2_Flambda_GS_def}
\frac{\ddroit}{\ddroit \upsilon}\mathcal{F}^{\lambda}[\upsilon,n^0]
&=&
\frac{\ddroit\mathcal{E}^\lambda_1(\upsilon)}{\ddroit \upsilon}-
\frac{\lambda}{1+C_u^2}=0,
\end{eqnarray}
which, according to Appendix~\ref{appendix:maxLengendreH2}, leads to the
{\it unique} solution
\begin{eqnarray}\label{upsilon=0}
\upsilon^\lambda=0,\hspace{0.4cm} 0\leq\lambda\leq1,
\end{eqnarray}
or, equivalently,
\begin{eqnarray}\label{vlambda_n0_unique}
 V^\lambda_g
- V^\lambda_u 
=
\langle 1\sigma_u^2 \vert
\hat{T}-\lambda[\hat{T}+\hat{V}_{\rm ne}]
\vert1\sigma_u^2\rangle
-\langle 1\sigma_g^2 \vert 
\hat{T}-\lambda[\hat{T}+\hat{V}_{\rm ne}]
\vert1\sigma_g^2\rangle
.
\end{eqnarray}
We thus conclude from Eq.~(\ref{Hlambdamatrix_2}) that the
ground-state AC can simply be constructed in the minimal basis when
multiplying the fully-interacting Hamiltonian by the interaction
strength $\lambda$: 
\begin{eqnarray}\label{Hlambdamatrix_GSAC}
\left[ \hat{H}^\lambda \right]=\lambda\left[
\hat{H}\right]+\lambda\mathcal{V}_u,
\end{eqnarray}
which means that the ground-state wavefunction does not vary along
the AC,
\begin{eqnarray}\label{Psi1lambda_GSAC}
\Psi_1^\lambda=\Psi_1, \hspace{0.4cm} 0\leq\lambda\leq1.
\end{eqnarray}
Note that the description in the minimal basis of the physical ground-state wavefunction of
H$_2$ becomes {\it
exact} in the dissociation limit. For dissociated systems, the Legendre--Fenchel
transform will however be
ill-defined in the sense that the functional
derivative of the energy with respect to the electron density does not
exist~\cite{TCA07_Ayers_nonlocality_paradox}. Let us therefore stress
that what is described here is the
near dissociation of H$_2$ when neglecting the overlap between $1s_A$
and $1s_B$ orbitals.

According to Eq.~(\ref{Psi1lambda_GSAC}), when approaching the dissociation 
limit, the exact value for the ground-state Hxc integrand
$\langle\Psi_1^\lambda\vert\hat{W}_{\rm ee}\vert\Psi_1^\lambda\rangle$
should therefore be expected to become independent on the interaction strength
$\lambda$. This was observed numerically
by Teale \etal~\cite{Teale:2009p2020} An important difference though between their calculations,
which were performed in large basis sets, and the analytical ones
presented here lies in the fact that, in the $\lambda$=0 limit, Teale
\etal~\cite{Teale:2009p2020}
obtain a single determinantal KS wavefunction while we need to use an
ensemble of two states to reproduce the ground-state density.
Consequently, the large increase in the integrand curvature
that Teale \etal~observed for large bond
distances 
when approaching the $\lambda$=0 limit 
cannot be
reproduced in the minimal basis. 

Let us finally mention that, by applying Nagy's formula in
Eq.~(\ref{nagy_ACHxcw})
for $w=0$ and using Eq.~(\ref{Psi1lambda_GSAC}), we obtain the following expression for the ground-state exchange--correlation energy  
\begin{eqnarray}\label{GSxcenergy_H2}
E_{\rm xc}[n^0]&=&\int^1_0
\ddroit {\lambda}\,
\langle\Psi_1^\lambda\vert\hat{W}_{\rm
ee}\vert\Psi_1^\lambda\rangle
-E_{\rm H}[n^0]
\nonumber\\
&=&
\langle\Psi_1\vert\hat{W}_{\rm
ee}\vert\Psi_1\rangle-E_{\rm H}[n^0],
\end{eqnarray}
which gives, in the dissociation
limit~\cite{H2minbasissetJChemEduc,Sharkas_JCP12_MCH},
\begin{equation}\label{GSxcdisslimitH2}
E_{\rm xc} [n^0] \underset{R \rightarrow +\infty}{\longrightarrow} -\frac{5}{8}.
\end{equation}  
Interestingly, since $E_g=E_u$ and $K={5}/{16}$ a.u. in the dissociation
limit~\cite{H2minbasissetJChemEduc,Sharkas_JCP12_MCH}, the first excitation energy $E_2-E_1$ of the physical system,
where
\begin{eqnarray}\label{1stXenergy}
E_2= \frac{1}{2} \left( {E}_g + {E}_u + \sqrt{\big({E}_g-{E}_u\big)^2 + 4 K^2} \right)
,
\end{eqnarray}
according to Eq.~(\ref{Hmatrix}),
reduces to
\begin{equation}\label{XEdisslimitH2}
E_2-E_1 \underset{R \rightarrow +\infty}{\longrightarrow} 2K=\frac{5}{8}.
\end{equation}  
Since the KS determinants are degenerate in the minimal basis, the non-interacting
KS excitation energy in Eq.~(\ref{eq:taylorexp1storder_GSdensity}) equals zero and the 
exchange--correlation DD energy computed for the ground-state
density becomes 
\begin{equation}\label{deltaxcn0limitH2}
\Delta_{\rm xc}^0[n^0]=E_2-E_1 \underset{R \rightarrow
+\infty}{\longrightarrow} -E_{\rm xc} [n^0].
\end{equation}  

\subsection{ACs for the ensemble}\label{subsubsec:theory:AC_ensemble_min_basis}

The AC constructed in Sec.~\ref{subsubsec:theory:AC_GS_min_basis} for the ground state of H$_2$ in a minimal
basis is also valid for the two-state ensemble $\{\Psi_1,\Psi_2\}^w$ where
\begin{eqnarray}\label{1stXwf}
\vert{\Psi}_2\rangle=\frac{1}{\sqrt{1+C_g^2}}\Big(C_g\vert1\sigma_g^2\rangle+\vert1\sigma_u^2\rangle\Big) 
\end{eqnarray}
is the physical excited state whose energy $E_2$ is given
in Eq.~(\ref{1stXenergy}) and 
\begin{eqnarray}\label{GSCg}
C_g=\frac{E_2-E_u}{K}.
\end{eqnarray}
Indeed, since the physical fully-interacting
Hamiltonian is simply scaled by the interaction strength along the
ground-state AC (see Eq.~(\ref{Hlambdamatrix_GSAC})),
both ground and excited states do not vary with $\lambda$, 
\begin{eqnarray}\label{Psiilambda_ensAC}
\Psi_i^\lambda=\Psi_i, \hspace{0.4cm} 0\leq\lambda\leq1, \hspace{0.4cm}
i=1,2,
\end{eqnarray}
and the density constraint of Nagy's AC is therefore fulfilled
\begin{eqnarray}\label{denconstraint_H2_nagy}
(1-w)\,n_{\Psi_1^{\lambda}}({\bf
r})+w\,n_{\Psi_2^{\lambda}}({\bf
r})
=
(1-w)\,n_{\Psi_1}({\bf
r})+w\,n_{\Psi_2}({\bf
r})
,\hspace{0.4cm} 0\leq\lambda\leq1.
\end{eqnarray}
Note that, due to the degeneracy of the non-interacting GOK states 
and according to Eq.~(\ref{eq:XE_Xe_DD}),
relation~(\ref{deltaxcn0limitH2}) remains fulfilled for any value of
$w$:
\begin{equation}\label{deltaxcwnwlimitH2}
\Delta_{\rm xc}^w[n^w]=E_2-E_1 \underset{R \rightarrow
+\infty}{\longrightarrow} -E_{\rm xc} [n^0].
\end{equation}  

Let us now discuss the construction of the GACE in the minimal basis. 
For simplicity, 
we will consider 
the situation where the local potential
that holds the ensemble density fixed, as both ensemble weight $\xi$ and
interaction strength $\lambda$ vary along the GACE, does not depend on $\xi$. In the particular case where $\lambda=1$, the fully-interacting
densities
of the ground and first-excited states should therefore be equal. 
Since, according to Eq.~(\ref{1stXwf}), the density of the excited state
can be expressed as
\begin{eqnarray}\label{1stXdens}
n_{{\Psi}_2}({\bf r})=\langle{\Psi}_2 \vert\hat{n}({\bf r})\vert{\Psi}_2\rangle
=
\frac{1}{1+C_g^2}\Big(C_g^2\,n_g({\bf r})+\,n_u({\bf r})\Big)
,
\end{eqnarray}
we deduce from Eq.~(\ref{GSdens}) the following condition 
\begin{eqnarray}\label{condCuCg}
C_u^2C_g^2-1=\Big(C_uC_g-1\Big)\Big(C_uC_g+1\Big)=0,
\end{eqnarray}
which, when combined with the inequalities $C_g<0$ and $C_u>0$, leads to 
\begin{eqnarray}\label{condCuCg_2}
C_uC_g=-1.
\end{eqnarray}
Since $K>0$, we finally conclude from Eqs.~(\ref{GSCu}) and
(\ref{GSCg}) that the $1\sigma_g^2$ and $1\sigma_u^2$
determinants should be degenerate in the fully-interacting system:
\begin{eqnarray}\label{condCuCg_3}
E_g=E_u.
\end{eqnarray}
Consequently, the fully-interacting Hamiltonian to be used in the GACE equals 
\begin{eqnarray}\label{Hmatrix_GACE}
\left[ \hat{H}^{1,\xi} \right] =
\left[
\begin{array}{c c}
E_g & K \\
K & E_g
\end{array}
\right].
\end{eqnarray}
The corresponding weight-independent ground-state wavefunction  
\begin{eqnarray}\label{GSwf_GACE}
\vert{\Psi_1^{1,\xi}}\rangle=
\vert\overline{\Psi}_1\rangle=
\frac{1}{\sqrt{2}}\Big(\vert1\sigma_g^2\rangle-\vert1\sigma_u^2\rangle\Big),
\end{eqnarray}
whose energy equals $\mathcal{E}_1^{1,\xi}=E_g-K$, describes the neutral
dissociated state of H$_2$ while the weight-independent excited state
\begin{eqnarray}\label{1stXwf_GACE}
\vert{\Psi_2^{1,\xi}}\rangle=
\vert\overline{\Psi}_2\rangle=
\frac{1}{\sqrt{2}}\Big(\vert1\sigma_g^2\rangle+\vert1\sigma_u^2\rangle\Big) 
,
\end{eqnarray}
whose energy equals $\mathcal{E}_2^{1,\xi}=E_g+K$, describes the ionic state.
It is then clear that the ensemble density remains fixed as the
ensemble weight $\xi$ varies: 
\begin{eqnarray}\label{ensdens_GACE}
n({\bf r})=(1-\xi)\,n_{\Psi_1^{1,\xi}}({\bf r})+\xi\,\,n_{\Psi_2^{1,\xi}}({\bf
r})
=\frac{1}{2}\Big(n_g({\bf r})+n_u({\bf r})\Big)
.
\end{eqnarray}

The GACE can now be constructed with the partially-interacting
Hamiltonian written in Eq.~(\ref{Hlambdamatrix}) by substituting the variables
$\overline{\mathcal{V}}_g$ and
$\overline{\mathcal{V}}_u$ for $V_g$ and $V_u$, respectively, with 
\begin{eqnarray}\label{new_variables_GACE}
{V}_i=\lambda E_g -
\langle 1\sigma_i^2 \vert \hat{T}+
\lambda\hat{W}_{\rm
ee}
\vert1\sigma_i^2\rangle +\lambda\overline{\mathcal{V}}_i, \hspace{0.4cm} i=g,u,
\end{eqnarray}
which leads to the following parameterization
\begin{eqnarray}\label{Hlambdamatrix_GACE}
\left[ \hat{H}^\lambda \right] =\lambda
\left[
\begin{array}{c c}
E_g+\overline{\upsilon}& K \\
K & 
E_g
\end{array}
\right]+\lambda\overline{\mathcal{V}}_u,
\end{eqnarray}
where
$\overline{\upsilon}=\overline{\mathcal{V}}_g-\overline{\mathcal{V}}_u$
is the parameter than defines uniquely (up to a constant) the local
potential in the minimal basis. The auxiliary ground- and excited-state energies
are therefore expressed as
\begin{eqnarray}\label{GSauxienergy_GACE}
\mathcal{E}^\lambda_1(\overline{\upsilon})= \frac{\lambda}{2} \left(
2{E}_g +\overline{\upsilon}
- \sqrt{\overline{\upsilon}^2 + 4 K^2}
\right)+\lambda\overline{\mathcal{V}}_u
,
\end{eqnarray}
and 
\begin{eqnarray}\label{1stauxienergy_GACE}
\mathcal{E}^\lambda_2(\overline{\upsilon})= \frac{\lambda}{2} \left(
2{E}_g +\overline{\upsilon}
+ \sqrt{\overline{\upsilon}^2 + 4 K^2}
\right)+\lambda\overline{\mathcal{V}}_u
,
\end{eqnarray}
respectively. According to Eq.~(\ref{Legrendre_Fenchel_Flambda_x}), we can thus express the Legendre--Fenchel transform for
the ensemble as
\begin{eqnarray}\label{Legrendre_Fenchel_Flambda_x_H2}
F^{\lambda,\xi}[n]=\underset{\overline{\upsilon}}{\rm sup}
\bigg\{
\mathcal{F}^{\lambda,\xi}[\overline{\upsilon},n]
\bigg\}
,
\end{eqnarray}
where, according to Eqs.~(\ref{VgVudef}) and (\ref{ensdens_GACE}), 
\begin{eqnarray}\label{Legrendre_Fenchel_Flambda_x_H2_2}
\mathcal{F}^{\lambda,\xi}[\overline{\upsilon},n]
&=&
(1-\xi)\mathcal{E}^\lambda_1(\overline{\upsilon})+\xi\,\mathcal{E}^\lambda_2(\overline{\upsilon})
-
\int {\rm d}{\bf
r}\,v({\bf r})\,{n}({\bf r}),
\nonumber\\
&=&\frac{\lambda}{2}\left(2{E}_g
+\overline{\upsilon}+(2\xi-1)\sqrt{\overline{\upsilon}^2 + 4
K^2}\right)+\lambda\overline{\mathcal{V}}_u
-\frac{1}{2}\Big( V_g+V_u\Big).
\end{eqnarray}
Since in our parameterization $\overline{\mathcal{V}}_u$ is a constant,
$V_u$ does not vary with $\overline{\upsilon}$ and $\ddroit
V_g/\ddroit \overline{\upsilon}=\lambda$, according to
Eq.~(\ref{new_variables_GACE}). Consequently, the maximizing
$\overline{\upsilon}^{\lambda,\xi}$ parameter in
Eq.~(\ref{Legrendre_Fenchel_Flambda_x_H2}) fulfills 
\begin{eqnarray}\label{deriv_Legrendre_Fenchel_Flambda_x_H2}
\frac{\ddroit}{\ddroit \overline{\upsilon}}\mathcal{F}^{\lambda,\xi}[\overline{\upsilon},n]
&=&
\frac{\lambda(2\xi-1)}{2}\frac{\overline{\upsilon}}{\sqrt{\overline{\upsilon}^2 + 4
K^2}}=0,
\end{eqnarray}
which leads to the {\it unique} solution 
\begin{eqnarray}\label{upsilonxlambda=0}
\overline{\upsilon}^{\lambda,\xi}=0,\hspace{0.4cm} 0\leq\lambda\leq1,
\hspace{0.4cm} 0\leq \xi\leq w,
\end{eqnarray}
or, equivalently,
\begin{eqnarray}\label{vlambdax_n_unique}
 V^{\lambda,\xi}_g
- V^{\lambda,\xi}_u 
=
\langle 1\sigma_u^2 \vert
\hat{T}+\lambda\hat{W}_{\rm ee}
\vert1\sigma_u^2\rangle
-\langle 1\sigma_g^2 \vert 
\hat{T}+\lambda\hat{W}_{\rm ee}
\vert1\sigma_g^2\rangle
.
\end{eqnarray}
We thus conclude from Eq.~(\ref{Hlambdamatrix_GACE}) that the GACE can be constructed in the minimal basis when using
the partially-interacting Hamiltonian
\begin{eqnarray}\label{Hlambdamatrix_GACE_final}
\left[ \hat{H}^{\lambda,\xi} \right] =\lambda
\left[
\begin{array}{c c}
E_g & K \\
K & E_g
\end{array}
\right]
+\lambda\overline{\mathcal{V}}_u.
\end{eqnarray}
In this simple model both ground- and excited-state
wavefunctions will therefore not vary along the GACE, 
\begin{eqnarray}\label{Psiilambda_GACE}
\Psi_i^{\lambda,\xi}=\overline{\Psi}_i, \hspace{0.4cm} 0\leq\lambda\leq1, \hspace{0.4cm}
0\leq \xi\leq w,\hspace{0.4cm} i=1,2,
\end{eqnarray}
and the auxiliary excitation energy equals
\begin{eqnarray}\label{auxXE_GACE}
\mathcal{E}_2^{\lambda,\xi}-\mathcal{E}_1^{\lambda,\xi}=2\lambda K.
\end{eqnarray}
According to Eq.~(\ref{eq:Exc_AC_wANDlambda_2}), the
ensemble exchange--correlation energy is then equal to 
\begin{eqnarray}\label{Excw_gace_H2}
{E}^w_{\rm xc}[n]={E}_{\rm xc}[n]+2K w.
\end{eqnarray}
Since the density $n$ defined in Eq.~(\ref{ensdens_GACE}) corresponds to
the exact ground-state density $n^0$ in the dissociation limit
of H$_2$, we obtain from Eqs.~(\ref{GSxcdisslimitH2}) and
(\ref{XEdisslimitH2}) 
\begin{eqnarray}\label{Excw_gace_H2_Rinfty}
{E}^w_{\rm xc}[n] \underset{R \rightarrow +\infty}{\longrightarrow}\Big(1-w\Big){E}_{\rm xc}[n]
,
\end{eqnarray}
or, equivalently,
\begin{eqnarray}\label{Deltaxcw_gace_H2_Rinfty}
{\Delta}^w_{\rm xc}[n] \underset{R \rightarrow
+\infty}{\longrightarrow}-{E}_{\rm xc}[n].
\end{eqnarray}

\subsection{
The GSxc approximation}\label{subsec:GSxcapprox}

From the ensemble exchange--correlation energy expression in
Eq.~(\ref{Excw_gace_H2_Rinfty}),
which is exact for the dissociated H$_2$ molecule in a minimal basis, 
we deduce the following DFA for a
two-state ensemble:
\begin{eqnarray}\label{GSxcfun_DFA}
{E}^{w,\mbox{\tiny DFA}}_{\rm xc}[n]=\Big(1-w\Big){E}^{\mbox{\tiny DFA}}_{\rm xc}[n]
,
\end{eqnarray}
or, equivalently,
\begin{eqnarray}\label{GSxcDelta_DFA}
{\Delta}^{w,\mbox{\tiny DFA}}_{\rm xc}[n]=-{E}^{\mbox{\tiny DFA}}_{\rm
xc}[n],
\end{eqnarray}
where any pure ground-state exchange--correlation density functional can in
principle be used. We thus define from
Eq.~(\ref{eq:taylorexp1storder_GSdensity}) the approximate {\it ground-state exchange--correlation
energy} (GSxc)-corrected excitation energy expression
\begin{eqnarray}\label{GSxcXE}
\Big(E_2-E_1\Big)_{\rm
GSxc}=\varepsilon^0_2-\varepsilon^0_1-{E}^{\mbox{\tiny DFA}}_{\rm
xc}[n^0],
\end{eqnarray}
where 
the
exchange--correlation energy computed for the ground-state density is 
subtracted from the KS orbital energy
difference.
Note
that in case of multiple excitations the latter  
will be replaced by a sum of orbital energy differences.

\section{Illustrative result: the $2^1\Sigma^+_g$ state of H$_2$ upon
bond stretching 
}\label{sec:discussion}

The first $^1\Sigma^+_g$ excitation energy in H$_2$ has
been computed within the GSxc approximation introduced in
Sec.~\ref{subsec:GSxcapprox}. Comparison is made with Full Configuration
Interaction (FCI) and regular TD-DFT results. 
The local density  (LDA)~\cite{dft-Vosko-CJP1980a} as well as
the semi-local Perdew--Burke--Ernzerhof (PBE)~\cite{dft-Perdew-PRL1996a} and 1994
Leeuwen--Baerends (LB94)~\cite{PRA94_Baerends_LB94} approximations have been considered.
The large aug-cc-pVQZ
basis set~\cite{augQZbe} has been used.
Calculations were performed with the DALTON2011
program~\cite{DALTON}.

\begin{figure}
\centering
{
\includegraphics[scale=0.5]{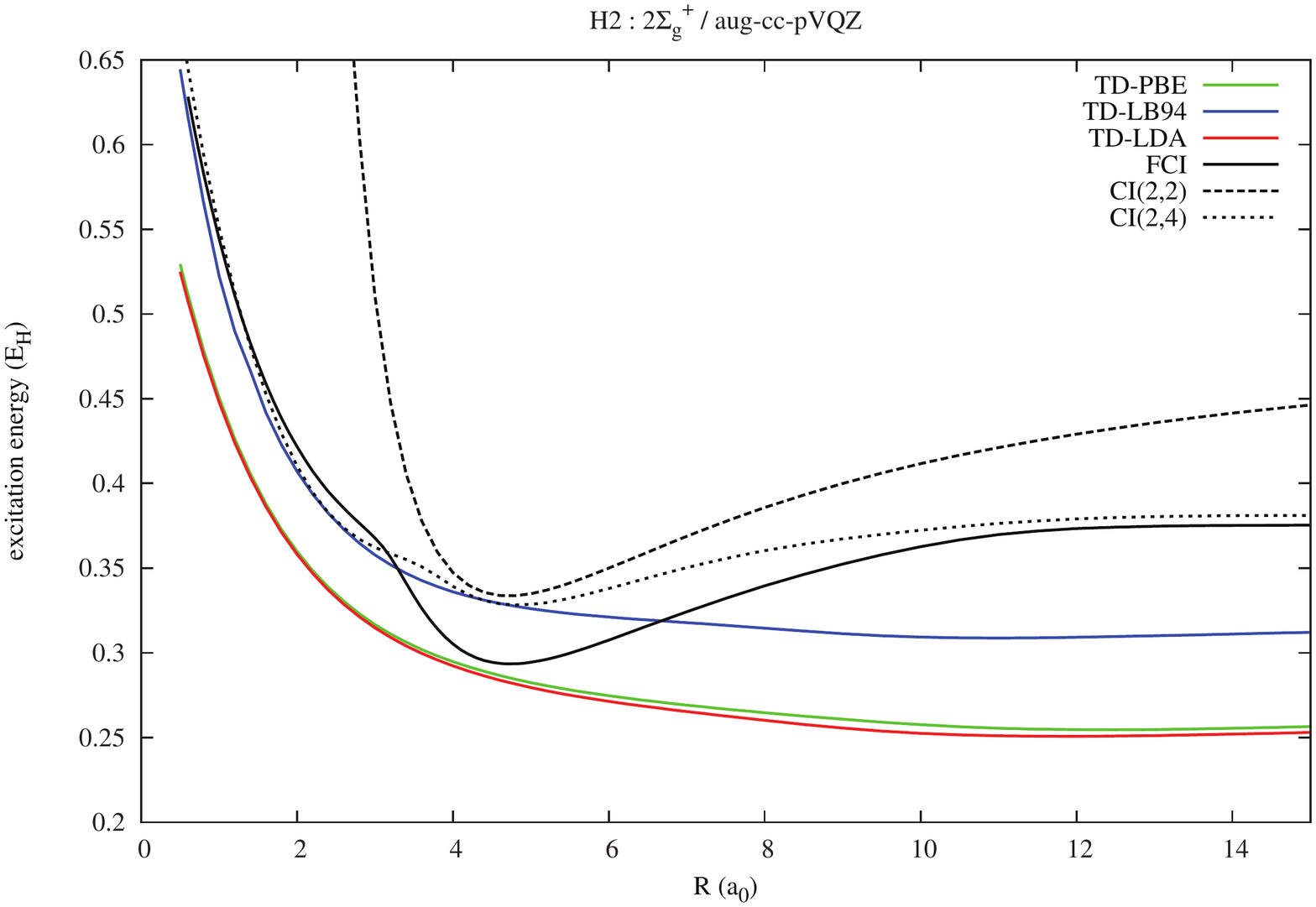}}
\caption{
First $^1\Sigma_g^+$ excitation energy in H$_2$ along the bond-breaking
coordinate obtained with regular TD-DFT (solid colored lines) and CI
(black lines)
methods. FCI (solid line) is compared to CI(2,2) (dashed line), where
the two electrons are distributed among the $1\sigma_g$ and $1\sigma_u$
orbitals, and to CI(2,4) (dotted line) where the two electrons are distributed among
the $1\sigma_g$, $2\sigma_g$, $1\sigma_u$ and $2\sigma_u$ orbitals.
}\label{fig:figure-intro}
\end{figure}

Regular adiabatic TD-DFT
fails in describing the $2^1\Sigma^+_g$ excited state of H$_2$ upon
bond stretching since, for bond distances larger than 3 a.u., this state exhibits a
strong doubly-excited character~\cite{JCP12_Baerends_tddmft}, as shown in
Fig.~\ref{fig:figure-intro}. The avoided crossing obtained at the FCI
level around $R$=3 a.u. indicates the change in character for the
$2^1\Sigma^+_g$ state, from singly
[$1\sigma_g\rightarrow 2\sigma_g$]
to doubly [$(1\sigma_g)^2\rightarrow(1\sigma_u)^2$] excited, while the TD-DFT curves remain associated
with the single excitation even for large bond distances. 
Before discussing the performance of the GSxc approximation, we should
first
stress that the minimal basis model on which it relies 
is exact for the ground $1^1\Sigma^+_g$ state of H$_2$ in the dissociation limit.
However, as shown by the
CI(2,2) and CI(2,4) excitation energy curves (see caption of
Fig.~\ref{fig:figure-intro}), it provides a qualitatively correct description of the
$2^1\Sigma^+_g$ state only in the range $4\leq R\leq 5$ a.u., where the
doubly-excited configuration $1\sigma_u^2$ is dominant in the
wavefunction. On the other hand, the singly-excited configuration $1\sigma_g
2\sigma_g$, which is not included into the minimal basis model, increasingly
dominates as $R$ decreases and becomes, for $R\geq5$ a.u., as important as the doubly-excited
configuration. In the latter case it
enables to describe the atomic $1s\rightarrow 2s$ excitation 
as $R\rightarrow+\infty$. The corresponding excitation
energy (3/8 a.u.) is indeed lower than the one associated with the
excitation from the neutral ground-state to the ionic dissociated state
(5/8 a.u.). The latter excitation is the only one described in the minimal
basis.
We should therefore not expect the GSxc
approximation to perform well for all bond distances when a large basis
set is used.

\begin{figure}
\centering
{
\includegraphics[scale=0.5]{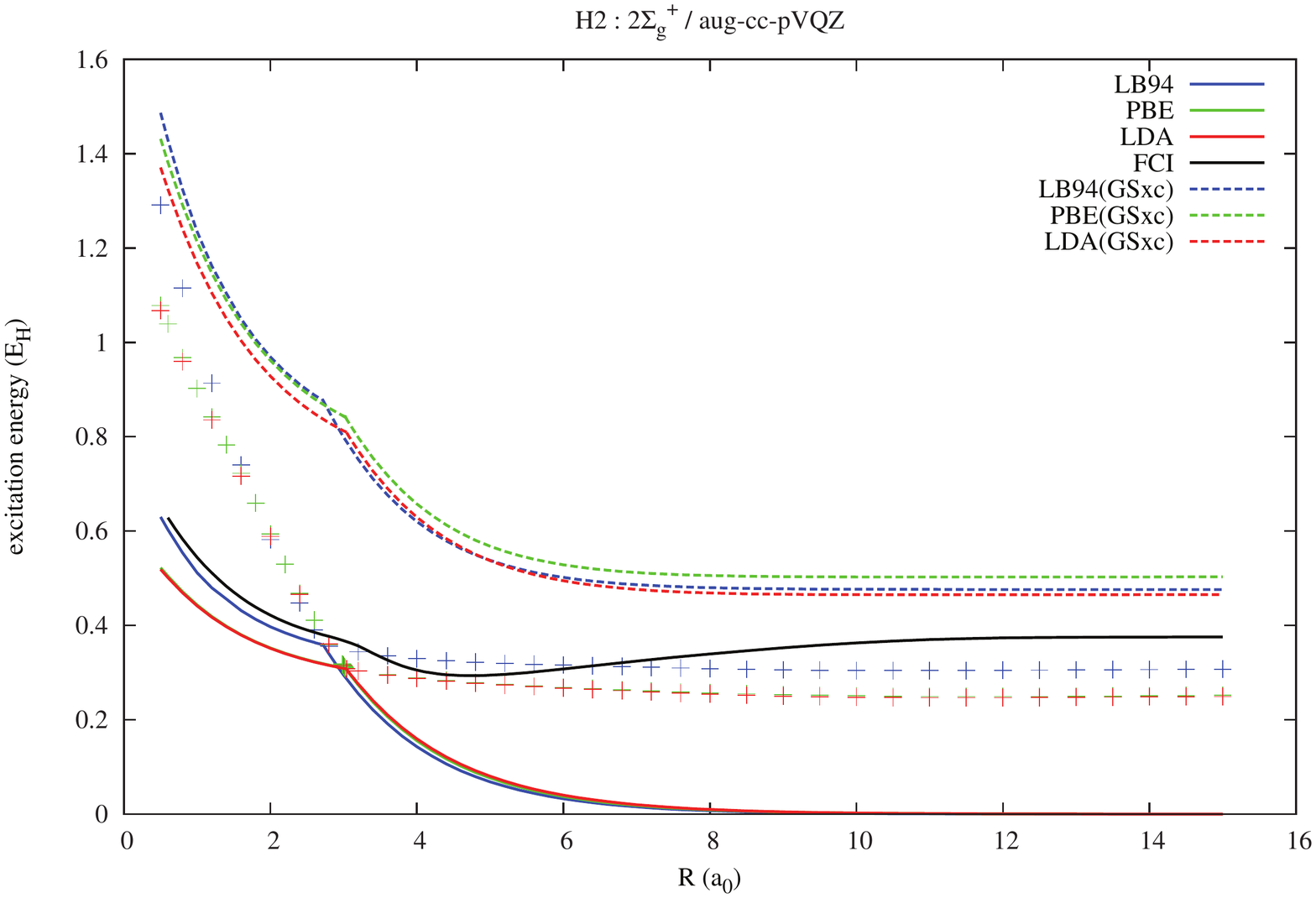}}
\caption{
First $^1\Sigma_g^+$ excitation energy in H$_2$ along the bond-breaking
coordinate obtained within the GSxc approximation 
with local and semi-local functionals (dashed colored lines).
Comparison is made with FCI (solid black line) and with the KS
excitation energy without the GSxc correction (solid colored lines).
Crossings of singly- and doubly-excited KS states are shown for each
functional with colored
"+" points. See
text for further details.
}\label{fig:figure-GSxc}
\end{figure}

We now discuss the results shown in Fig.~\ref{fig:figure-GSxc}. 
Let us first stress that using a two-state ensemble enables the description
of the
double excitation $(1\sigma_g)^2\rightarrow(1\sigma_u)^2$ upon bond
stretching, as reflected by the sudden change in slope for the excitation
energy curves, even when the GSxc correction is not employed. In the latter case the
computed excitation energy simply equals the KS orbital energy difference
$\varepsilon^0_{2\sigma_g}-\varepsilon^0_{1\sigma_g}$ when $R\leq R_c$ and
$2(\varepsilon^0_{1\sigma_u}-\varepsilon^0_{1\sigma_g})$ when
$R\geq R_c$, where $R_c$ denotes the distance for which the crossing
between the singly-excited $1\sigma_g2\sigma_g$ and doubly-excited
$1\sigma_u^2$ KS
states occurs. 
Interestingly, in the particular case of LB94, this crossing is
relatively close to the FCI avoided crossing ($R_c\approx 3$
a.u.). A slightly larger $R_c$ value is obtained with LDA and PBE and,
for $R\leq R_c$, the computed excitation energies are less accurate relative to
LB94. This was expected as the latter approximation 
includes corrections for a proper description of the
exchange--correlation potential in the asymptotic region of 
atoms~\cite{PRA94_Baerends_LB94}. For $R\geq R_c$, the excitation energy decreases rapidly
to zero
with the bond distance for all the functionals simply
because the $1\sigma_g$ and $1\sigma_u$ KS orbitals or, equivalently,
the $1\sigma_g^2$ and $1\sigma_u^2$ KS determinants become degenerate,
like in the minimal basis. As shown in Fig.~\ref{fig:figure-GSxc}
employing the
GSxc correction enables to recover reasonable excitation energies in the 
dissociation limit, with a slight overestimation relative to FCI though. This is not too
surprising since, as mentioned   
previously, the {\it neutral} $\rightarrow$ {\it ionic} excitation underlying the GSxc
approximation is higher than the atomic $1s\rightarrow 2s$ excitation.     
On the other hand, for shorter bond distances, the GSxc-corrected excitation energies are
much too high. In the range $4\leq R\leq 5$ a.u., the error
is partially due to the fact that, in the minimal basis model, the KS
determinants are degenerate while, in the larger aug-cc-pVQZ basis, they
are not. The large error at equilibrium ($R=1.4$ a.u.) is due to the absence of
single excitations in the minimal basis model. 
Obviously the singly excited
$1\sigma_g2\sigma_g$ configuration should be included into the ensemble
in order to improve the GSxc model, especially in that region. 
As it might be difficult to reproduce the FCI avoided crossing without
treating explicitly couplings between the states included into the ensemble,
the development of a multi-determinant GOK-DFT scheme is an appealing
alternative. 
Pastorczak \etal~\cite{PRA13_Pernal_srEDFT}~recently proposed such an approach based on the range
separation of the two-electron repulsion. As discussed briefly in
Sec.~\ref{sec:EsrDFT},
a range-dependent GACE could be used in this context for the development
of appropriate short-range ensemble exchange--correlation density functionals. 

\section{Perspective: range-dependent GACE 
}\label{sec:EsrDFT}

Pastorczak \etal~\cite{PRA13_Pernal_srEDFT} recently formulated a multi-determinant extension
of GOK-DFT that relies on the separation of the two-electron
repulsion into long-range (lr) and short-range (sr) parts 
\begin{eqnarray}\label{ee=lr+sr}
\frac{1}{r_{12}}=w^{\rm lr,\mu}_{\rm ee}(r_{12})+w^{\rm sr,\mu}_{\rm
ee}(r_{12}),
\end{eqnarray}
where $\mu$ is a parameter that controls the range separation with $w^{\rm lr,\mu}_{\rm ee}(r_{12})=1/r_{12}$ in the
$\mu\rightarrow+\infty$ limit and $w^{\rm lr,\mu}_{\rm
ee}(r_{12})=0$ for $\mu=0$. 
By analogy with ground-state multi-determinant range-separated
DFT~\cite{erferfgaufunc}, they decomposed the universal GOK functional as follows 
\begin{eqnarray}\label{fwsplitrange}
F^w[n]&=&F^{{\rm lr,\mu},w}[n]+{E}^{{\rm sr,\mu},w}_{\rm Hxc}[n],
\end{eqnarray}
where the universal long-range GOK functional is defined as
\begin{eqnarray}\label{Flrwdef}
F^{{\rm lr,\mu},w}[n]&=&\underset{\{\Psi_1,\Psi_2\}^w\rightarrow n}{\rm min}\Big\{
(1-w)\,\langle\Psi_1\vert\hat{T}+\hat{W}^{\rm lr,\mu}_{\rm ee}\vert\Psi_1\rangle
+w\,\langle\Psi_2\vert\hat{T}+\hat{W}^{\rm lr,\mu}_{\rm
ee}\vert\Psi_2\rangle\Big\}
\nonumber\\
&=&
(1-w)\,\langle\Psi^{\mu,w}_1\vert\hat{T}+\hat{W}^{\rm lr,\mu}_{\rm
ee}\vert\Psi^{\mu,w}_1\rangle
+w\,\langle\Psi^{\mu,w}_2\vert\hat{T}+\hat{W}^{\rm lr,\mu}_{\rm
ee}\vert\Psi^{\mu,w}_2\rangle,
\end{eqnarray}
and ${E}^{{\rm sr,\mu},w}_{\rm Hxc}[n]$ is the $\mu$-dependent
complementary short-range Hxc density functional for the ensemble.
According to the GOK variational principle in Eq.~(\ref{ensenergyvarprinc}),
the exact ensemble energy can then be written as follows
\begin{eqnarray}\label{Esrdftenergy}
E^w&=&
(1-w)\,\langle\tilde{\Psi}^{\mu,w}_1
\vert\hat{T}+\hat{W}^{\rm lr,\mu}_{\rm ee}\vert\tilde{\Psi}^{\mu,w}_1\rangle
+w\,\langle\tilde{\Psi}^{\mu,w}_2
\vert\hat{T}+\hat{W}^{\rm lr,\mu}_{\rm
ee}\vert\tilde{\Psi}^{\mu,w}_2\rangle
\nonumber\\
&&+{E}^{{\rm sr,\mu},w}_{\rm Hxc}[n^w]
+\int \ddroit{\bf r}\,v_{\rm
ne}({\bf r})\,n^w({\bf r}),
\end{eqnarray}
where the auxiliary long-range-interacting wave functions
$\tilde{\Psi}^{\mu,w}_i\;(i=1,2)$ that reproduce the exact ensemble
density $n^w$ fulfill the following
self-consistent equations:
\begin{eqnarray}\label{sceqensemblesrdft}
&&\Big(\hat{T}+\hat{W}^{\rm lr,\mu}_{\rm ee}+\hat{V}_{\rm
ne}+\hat{{V}}^{{\rm sr,\mu},w}_{\rm
Hxc}[n^w]\Big)\vert\tilde{\Psi}^{\mu,w}_i\rangle=\tilde{\mathcal{E}}^{\mu,w}_{i}\vert\tilde{\Psi}^{\mu,w}_i\rangle,
\hspace{0.4cm} i=1,2,
\nonumber\\
\nonumber\\
&&{\displaystyle
\hat{{V}}^{{\rm sr,\mu},w}_{\rm Hxc}[n]=
\int \ddroit{\mathbf r}\,\frac{\delta {E}^{{\rm sr,\mu},w}_{\rm Hxc}}{\delta n({\bf
r})}[n]\,\hat{n}({\bf r})
.   
}
\end{eqnarray}
While regular GOK-DFT and wavefunction theory approaches are recovered
in the $\mu=0$ and $\mu\rightarrow+\infty$ limits, respectively,  
an {\it exact} 
state-average multi-determinant DFT is obtained for $0<\mu<+\infty$.

For convenience, Pastorczak \etal~\cite{PRA13_Pernal_srEDFT} substituted the ground-state
short-range Hxc functional ${E}^{{\rm sr,\mu},0}_{\rm Hxc}[n]={E}^{{\rm
sr,\mu}}_{\rm Hxc}[n]$ for the ensemble one in their practical
calculations. This is a crude approximation which obviously can
 have an impact on the accuracy of the computed excitation energy,
 especially if small $\mu$ values are used~\cite{JCPunivmu,fromager2013}, since the range-separated
 approach is then closer to GOK-DFT than wavefunction theory. Better
 approximations might be developed from a range-dependent GACE. For that
 purpose we 
 introduce the auxiliary equations
\begin{eqnarray}\label{nlacw}
\Big(\hat{T}+\hat{W}^{\rm lr,\nu}_{\rm ee}+\hat{V}^{\nu,\xi}
\Big)\vert\Psi^{\nu,\xi}_i\rangle=\mathcal{E}_i^{\nu,\xi}\vert\Psi^{\nu,\xi}_i\rangle,
\hspace{0.2cm} i=1,2,
\end{eqnarray}
where the local potential 
$\hat{V}^{\nu,\xi}=\int \ddroit{\bf r}\,v^{\nu,\xi}({\bf r})\,\hat{n}({\bf r})$
ensures that the density constraint
\begin{eqnarray}\label{nlACwdensconstraints}
n({\bf r})&=&(1-\xi)\,n_{\Psi_1^{\nu,\xi}}({\bf
r})+\xi\,n_{\Psi_2^{\nu,\xi}}({\bf
r}),
\hspace{0.4cm} 0\leq \nu< +\infty,
\hspace{0.4cm} 0\leq \xi\leq w, 
\end{eqnarray}
is fulfilled.
By integration of the universal long-range GOK functional over the interval $[\mu,+\infty[$ 
we obtain from Eqs.~(\ref{fwsplitrange}), (\ref{Flrwdef}), (\ref{nlacw})
and (\ref{nlACwdensconstraints}),
\begin{eqnarray}\label{nlACsrHxcw}
\displaystyle
E^{{\rm sr,\mu},w}_{\rm Hxc}[n]&=&
\int^{+\infty}_\mu
\ddroit\nu\,\frac{\ddroit}{\ddroit \nu}F^{{\rm lr},\nu,w}[n]
\nonumber\\
&=&
(1-w)\,\int^{+\infty}_\mu
\ddroit\nu\,
\frac{\ddroit\mathcal{E}_1^{\nu,w}}{\ddroit\nu}
+w\,\int^{+\infty}_\mu
\ddroit\nu\,
\frac{\ddroit\mathcal{E}_2^{\nu,w}}{\ddroit\nu}
\nonumber
\\
&&-
 \int^{+\infty}_\mu\ddroit\nu\int \ddroit\mathbf{r}\,
\frac{\partial v^{\nu,w}({\bf
r})}{\partial \nu}\,n({\bf r})
,
\end{eqnarray}
which leads, according to the Hellmann--Feynman theorem, to the final expression
\begin{eqnarray}\label{nlACsrHxcw_2}
\displaystyle
E^{{\rm sr,\mu},w}_{\rm Hxc}[n]&=&
(1-w)
\int^{+\infty}_\mu
\ddroit\nu\,
\langle\Psi_1^{\nu,w}\vert\frac{\partial\hat{W}^{\rm lr,\nu}_{\rm
ee}}{\partial \nu}\vert\Psi_1^{\nu,w}\rangle
\nonumber\\
&&
+w\int^{+\infty}_\mu
\ddroit\nu\,\langle\Psi_2^{\nu,w}\vert\frac{\partial\hat{W}^{\rm lr,\nu}_{\rm
ee}}{\partial\nu}\vert\Psi_2^{\nu,w}\rangle.
\end{eqnarray}
By analogy with GOK-DFT, we use a weight-independent definition for the ensemble short-range Hartree
density-functional energy, 
\begin{align} \label{srDFTfunHdef}
E^{{\rm sr},\mu,w}_{\rm H}[n]=E^{{\rm sr},\mu}_{\rm H}[n]= \frac{1}{2}\int\int {\rm d}{\mathbf{r}}{\rm d}{\mathbf{r'}}n(\mathbf{r})n(\mathbf{r'})w^{\rm sr,\mu}_{\rm ee}\left(\vert {\bf r}-{\bf r'} \vert\right),
\end{align}
and thus define the short-range exchange--correlation energy for
the ensemble as
\begin{align} \label{srDFTfunxcHdef}
E^{{\rm sr,\mu},w}_{\rm xc}[n]=E^{{\rm sr,\mu},w}_{\rm Hxc}[n]-E^{{\rm
sr},\mu}_{\rm H}[n].
\end{align}
Like in the linear GACE that was introduced in
Sec.~\ref{subsec:theory:ACensemble}, the exact deviation of the ensemble short-range
exchange--correlation energy from the ground-state one can be derived by
integration over
the ensemble weight:
\begin{eqnarray}\label{eq:Exc_AC_wANDlambda_sr}
{E}^{{\rm sr},\mu,w}_{\rm xc}[n]&=&
{E}^{{\rm sr},\mu}_{\rm xc}[n]+
\int_0^w
\ddroit \xi\, \Delta^{{\rm sr},\mu,\xi}_{\rm xc}[n]
,
\end{eqnarray}
where
\begin{eqnarray}\label{eq:Exc_AC_wANDlambda_sr_2}
\Delta^{{\rm sr},\mu,\xi}_{\rm xc}[n]
&=&\frac{\ddroit E^{{\rm sr,\mu},\xi}_{\rm Hxc}[n] }{ \ddroit \xi} 
\nonumber\\
&=&\int^{+\infty}_\mu\ddroit\nu\,\frac{\ddroit^2 F^{{\rm lr,\nu},\xi}[n]}{\ddroit \nu \ddroit \xi} 
,
\end{eqnarray}
will be referred to as the {\it short-range 
exchange--correlation DD}
since it reduces to the standard exchange--correlation DD when $\mu=0$.
By analogy with the linear GACE (see
Appendix~\ref{appendix:deriv_Flambdax}), the derivative of
the long-range GOK functional with respect to the ensemble weight equals
\begin{eqnarray}\label{dFmux}
\displaystyle 
\frac{\ddroit F^{{\rm lr,\nu},\xi}[n]}{\ddroit
\xi}&=&\mathcal{E}^{\nu,\xi}_2-\mathcal{E}^{\nu,\xi}_1,
\end{eqnarray}
which leads to 
\begin{eqnarray}\label{dFmux_diffmuinfty}
\displaystyle 
\Delta^{{\rm sr},\mu,\xi}_{\rm xc}[n]
&=&
\Big(\mathcal{E}^{+\infty,\xi}_2-\mathcal{E}^{+\infty,\xi}_1\Big)
-
\Big(\mathcal{E}^{\mu,\xi}_2-\mathcal{E}^{\mu,\xi}_1\Big)
.
\end{eqnarray}
In the particular case where $\xi=w$ and $n$ equals the exact ensemble
density $n^w$, the first term 
on the right-hand side of Eq.~(\ref{dFmux_diffmuinfty}) becomes the excitation energy $E_2-E_1$
of the true physical system
while the second term reduces to the excitation energy
$\tilde{\mathcal{E}}^{\mu,w}_2-\tilde{\mathcal{E}}^{\mu,w}_1 $ of the
long-range-interacting
system whose ensemble density equals $n^w$ (see
Eq.~(\ref{sceqensemblesrdft})), leading thus to the exact expression 
\begin{eqnarray}\label{XEenssrdft}
\displaystyle 
E_2-E_1&=&
\tilde{\mathcal{E}}^{\mu,w}_2-\tilde{\mathcal{E}}^{\mu,w}_1
+\Delta^{{\rm sr},\mu,w}_{\rm xc}[n^w].
\end{eqnarray}
As readily seen from Eqs.~(\ref{eq:Exc_AC_wANDlambda_sr_2}) and
(\ref{XEenssrdft}), neglecting the weight dependence of the ensemble
short-range exchange--correlation functional is
equivalent to approximating the excitation energy with the
long-range interacting one. In order to investigate the variation in
$w$ and $\mu$ of the short-range exchange--correlation DD contribution,
a simple procedure would consist in neglecting the weight dependence in
the ensemble short-range exchange--correlation density-functional potential
as Pastorczak \etal~\cite{PRA13_Pernal_srEDFT} did in their range-separated ensemble calculations,
and computing the excitation energy difference
$(E_2-E_1)-(\tilde{\mathcal{E}}^{\mu,w}_2-\tilde{\mathcal{E}}^{\mu,w}_1)$
at the CI level for various systems. 
The derivation of exact Taylor expansions in $w$ and $\mu$ for the short-range
exchange--correlation DD, in the light of
Sec.~\ref{subsec:theory:ACensemble} and Ref.~\cite{erferfgaufunc}, would
also be of interest for the development of approximate short-range
ensemble
functionals. Work is currently in progress in these directions.

\section{Conclusions}\label{sec:conclusions}

A generalized adiabatic connection for ensembles (GACE) has been presented in
this work. In contrast to the adiabatic connection (AC)
proposed initially by Nagy~\cite{Nagy_ensAC}, both ensemble weights and interaction strength
vary along the GACE while the ensemble density is held fixed. For clarity the
theory has been presented for non-degenerate two-state ensembles but 
the 
GACE can in principle be constructed for any ensemble consisting of
an arbitrary number of non-degenerate states and complete sets of
degenerate states~\cite{PRA_GOK_RRprinc}. Within such a
formalism an exact expression for the deviation of the ensemble
exchange--correlation density-functional energy from the conventional
ground-state one has been 
derived. Levy's stringent constraint of Ref.~\cite{PRA_Levy_XE-N-N-1} has
been recovered
when expanding the ensemble exchange--correlation functional through second
order in the ensemble weight. In addition, an explicit expression for the
exchange--correlation derivative discontinuity contribution to this
condition has been obtained
within the GACE. In the light of the recent work of
Teale~\etal~\cite{Teale:2009p2020,Teale:2010,Teale:2010b}
on the accurate computation of ground-state ACs, we briefly explained how
the GACE could be constructed by using a Legendre--Fenchel
transform for ensembles.
As an illustration, the GACE has been derived analytically for the H$_2$
model system in a minimal basis, providing thus a simple
density-functional approximation for two-state ensembles. This approximation has
been tested with a large basis on the calculation of the first
$^1\Sigma^+_g$ excitation energy in H$_2$ upon bond stretching. Encouraging results were obtained
at large distance (the double excitation could be described) but better 
ensemble exchange--correlation functionals are needed for describing the
excitation at all bond distances, especially in order to reproduce the
avoided crossing at $R=3$ a.u. A more accurate description of the
GACE would be useful for developing such functionals. 
Following Pastorczak \etal~\cite{PRA13_Pernal_srEDFT}, we finally discussed as a perspective the development of a
state-average multi-determinant DFT approach based on a range-dependent GACE.
Exact expressions for the complementary short-range ensemble
exchange--correlation density-functional energy have been derived and
guidelines for the development of density-functional approximations have
been provided. Work is currently in progress in this direction. We
hope that the paper will stimulate further developments in ensemble DFT.      

\section*{Acknowledgments}

E.F thanks Andrew Teale, Andreas Savin, Trygve Helgaker, Stefan Knecht, Julien Toulouse
and Alex Borgoo for fruitful discussions. The authors would like to
thank the reviewers for their numerous comments, especially on the
$v$-representability problem, the use of Legendre transforms in
nearly dissociated systems and for suggesting to use imaginary
temperatures in Boltzmann factors in order to connect GOK-DFT with TD-DFT.
Such a connection should obviously be investigated further in the future.
\appendices

\section{Derivative of the partially-interacting GOK
functional with respect to the ensemble weight}\label{appendix:deriv_Flambdax}

When rewriting, according to Eqs.~(\ref{acw}) and
(\ref{ACwdensconstraints}), the partially-interacting GOK functional as
\begin{eqnarray}\label{Fxlambda_auxenergies}
F^{\lambda,\xi}[n]&=&
(1-\xi)\,\mathcal{E}_1^{\lambda,\xi}
+\xi\,\mathcal{E}_2^{\lambda,\xi}-\int \ddroit{\bf r}\,v^{\lambda,\xi}({\bf
r})\,{n}({\bf r})
\end{eqnarray}
we obtain 
\begin{eqnarray}\label{dFlambax_proof}
\displaystyle 
\frac{\ddroit F^{\lambda,\xi}[n]}{\ddroit
\xi}&=&\mathcal{E}^{\lambda,\xi}_2-\mathcal{E}^{\lambda,\xi}_1
+(1-\xi)\,\frac{\ddroit\mathcal{E}_1^{\lambda,\xi}}{\ddroit
\xi}+\xi\,\frac{\ddroit\mathcal{E}_2^{\lambda,\xi}}{ \ddroit \xi}
\nonumber\\
&&
-
\int \ddroit{\bf r}\,\frac{\partial v^{\lambda,\xi}({\bf r})}{\partial \xi}\,{n}({\bf r}),
\end{eqnarray}
which, according to the Hellmann--Feynman theorem in
Eq.~(\ref{eq:HFthweight}) and the
density constraint in Eq.~(\ref{ACwdensconstraints}), leads to
Eq.~(\ref{dFlambax}).

\section{Exact local potential for the non-interacting
ensemble}\label{appendix:GOKpot}

According to the GOK variational principle the density $n$ for which the
GACE is constructed minimizes the density-functional ensemble energy  
\begin{eqnarray}\label{ensembleDFTenerv1x}
\mathcal{E}^{\xi}[\rho]=
T^{\xi}_{\rm s}[\rho]+{E}^{\xi}_{\rm Hxc}[\rho]
+\int \ddroit{\bf r}\,\bigg(v^{1,\xi}
({\bf r})+C\bigg)\,\rho({\bf r})
\end{eqnarray}
where $C$ is an arbitrary constant.
The minimum equals
$(1-\xi)\mathcal{E}_1^{1,\xi}+\xi\,\mathcal{E}_2^{1,\xi}+CN$ where $N$
denotes the number of electrons (which is fixed in this work). Consequently
\begin{eqnarray}\label{diffensembleDFTenerv1x}
&&\left.\frac{\delta
}{\delta \rho({\bf r})}
\Bigg[\mathcal{E}^{\xi}[\rho]
+\mu^\xi\Bigg(\int \ddroit{\mathbf{r}\,\rho(\mathbf{r})} -N\Bigg)
\Bigg]
\right
|
_{\rho=n}
\nonumber\\
&&=\frac{\delta T^{\xi}_{\rm s}}{\delta \rho({\bf r})}[n]+
\frac{\delta {E}^{\xi}_{\rm Hxc}}{\delta \rho({\bf r})}[n]+v^{1,\xi}({\bf
r})+C+\mu^\xi
\nonumber
\\
&&=0,
\end{eqnarray}
where the Lagrange multiplier $\mu^\xi$ is the chemical potential. When
choosing $C=-\mu^\xi$,
we finally obtain Eq.~(\ref{eq:vreal_KS_connection}) since
\begin{eqnarray}\label{diffTsx}
\frac{\delta T^{\xi}_{\rm s}}{\delta \rho({\bf r})}[n]=-v^{0,\xi}({\bf
r}).
\end{eqnarray}
\section{Maximum of the ground-state Legendre--Fenchel transform for
H$_2$ in a minimal basis}\label{appendix:maxLengendreH2}
According to Eq.~(\ref{GSauxienergy})
the first-order derivative of the auxiliary ground-state energy can be expressed as
\begin{eqnarray}\label{MaxLegrendre_Fenchel_H2_Flambda_GS_def_proof}
\frac{\ddroit\mathcal{E}^\lambda_1(\upsilon)}{\ddroit \upsilon}
=
\frac{\lambda}{2} \left(1
- \frac{\delta+\upsilon}{\sqrt{\big(\delta+\upsilon\big)^2 + 4 K^2}}
\right)
,
\end{eqnarray}
where $\delta=E_g-E_u$. Using
\begin{eqnarray}\label{a+ba-b}
\left[\sqrt{\big(\delta+\upsilon\big)^2 + 4 K^2}-(\delta+\upsilon)\right]
\left[\sqrt{\big(\delta+\upsilon\big)^2 + 4
K^2}+(\delta+\upsilon)\right]
=4K^2,
\end{eqnarray}
Eq.~(\ref{MaxLegrendre_Fenchel_H2_Flambda_GS_def_proof}) becomes
\begin{eqnarray}\label{MaxLegrendre_Fenchel_H2_Flambda_GS_def_proof_2}
\frac{\ddroit\mathcal{E}^\lambda_1(\upsilon)}{\ddroit \upsilon}
&=&
\frac{2\lambda K^2}{\sqrt{\big(\delta+\upsilon\big)^2 + 4 K^2}
\left[\sqrt{\big(\delta+\upsilon\big)^2 + 4
K^2}+(\delta+\upsilon)\right]
}
\nonumber\\
&=&
\frac{2\lambda
K^2}{(\delta+\upsilon)\left[(\delta+\upsilon)+\sqrt{\big(\delta+\upsilon\big)^2 + 4
K^2}\right]
+ 4
K^2}
.
\end{eqnarray}
Since, according to Eqs.~(\ref{GSenergy}) and (\ref{GSCu}), 
\begin{eqnarray}\label{Cudelta}
1+C^2_u&=&1+\frac{1}{4K^2}\left(\delta+\sqrt{\delta^2 + 4
K^2}\right)^2
\nonumber\\
&=&\frac{4K^2+\delta\left(\delta+\sqrt{\delta^2 + 4K^2}\right)}{2K^2},
\end{eqnarray}
we conclude that Eq.~(\ref{MaxLegrendre_Fenchel_H2_Flambda_GS_def}) is equivalent to
\begin{eqnarray}\label{fupsilon=fzero}
f(\upsilon)=f(0),
\end{eqnarray}
where the function $f$ is defined as
\begin{eqnarray}\label{ffunction}
f(\upsilon)=(\delta+\upsilon)\left[(\delta+\upsilon)+\sqrt{\big(\delta+\upsilon\big)^2 + 4
K^2}\right]
.
\end{eqnarray}
Finally, since 
\begin{eqnarray}\label{derivffunction}
\frac{\ddroit f}{\ddroit\upsilon}
=\frac{
\left[(\delta+\upsilon)+
\sqrt{\big(\delta+\upsilon\big)^2 + 4
K^2}
\right]^2
}
{
\sqrt{\big(\delta+\upsilon\big)^2 + 4
K^2}
}>0
,
\end{eqnarray}
$f$ is monotonically increasing with $\upsilon$ which leads to Eq.~(\ref{upsilon=0}).

\bibliographystyle{tMPH}


\label{lastpage}

\end{document}